\documentclass[aps,pra,twocolumn,nofootinbib,reprint,showpacs]{revtex4-1}

\usepackage{amssymb}
\usepackage{graphicx}
\usepackage{amsmath}
\usepackage{braket}
\usepackage{hyperref}
\usepackage{bbold}
\usepackage{xcolor}
\usepackage{enumerate}

\DeclareMathOperator*{\argmax}{arg\,max}

\usepackage{environ}
\makeatletter
\NewEnviron{Lalign}{\tagsleft@true\begin{align}\BODY\end{align}}
\makeatother

\def\<{\langle}

\def\H{ {\cal H} }

\def\I{ \mathbb{1} }

\def\p{\boldsymbol{p}}

\begin{document}

\title{Quasi-autonomous quantum thermal machines and quantum to classical energy flow}

\author{Max F. Frenzel}
\affiliation{Controlled Quantum Dynamics Theory Group, Imperial College London, Prince Consort Road, London SW7 2BW, UK}
\author{David Jennings}
\affiliation{Controlled Quantum Dynamics Theory Group, Imperial College London, Prince Consort Road, London SW7 2BW, UK}
\author{Terry Rudolph}
\affiliation{Controlled Quantum Dynamics Theory Group, Imperial College London, Prince Consort Road, London SW7 2BW, UK}
\date{\today}

\begin{abstract}
There are both practical and foundational motivations to consider the thermodynamics of quantum systems at small scales. Here we address the issue of autonomous quantum thermal machines that are tailored to achieve some specific thermodynamic primitive, such as work extraction in the presence of a thermal environment, while having minimal or no control from the macroscopic regime. Beyond experimental implementations, this provides an arena in which to address certain foundational aspects such as the role of coherence in thermodynamics, the use of clock degrees of freedom and the simulation of local time-dependent Hamiltonians in a particular quantum subsystem. For small-scale systems additional issues arise. Firstly, it is not clear to what degree genuine ordered thermodynamic work has been extracted, and secondly non-trivial back-actions on the thermal machine must be accounted for. We find that both these aspects can be resolved through a judicious choice of quantum measurements that magnify thermodynamic properties up the ladder of length-scales, while simultaneously stabilizing the quantum thermal machine. Within this framework we show that thermodynamic reversibility is obtained in a particular Zeno limit, and finally illustrate these concepts with a concrete example involving spin-systems.

\end{abstract}


\maketitle

\section{Introduction\label{sec:Introduction}}

The issue of work in arbitrary-scale quantum systems turns out to be quite subtle, and a good deal of recent studies \cite{Brandao2011,Brandao2013, Skrzypczyk2013,Linden2010,Brunner2014,Navascues2014, Gallego2015,Roncaglia2014,Aberg2011,Horodecki2011,Dahlsten2011a,Frenzel2014d, Sagawa2008,Toyabe2010,Ashida2014,Skrzypczyk2014b,Anders,Jarzynski2015,Gour2013a, Ng2014b,Nicolas2015,Wehner2015} have analysed varying notions of work for finite-sized quantum systems. More recently the role that quantum-mechanical properties, such as coherence, play in work extraction are being addressed using resource-theoretic formulations. For example, it has been pointed out in \cite{Lostaglio2014} that free energies do not constitute proper coherence measures, and so to properly quantify the thermodynamic value of quantum coherence it is necessary to develop measures that go beyond free energies. In subsequent analysis \cite{Oppenheim2014aa,Lostaglio2015}, general upper and lower-bounds have been developed to constrain such coherent transformations under very general thermodynamic operations. Moreover, in \cite{Aberg2014,Korzekwa} the question of work extraction from an arbitrary qubit state has been analysed in a context which explicitly models the coherence resources that are required to extract work from the qubit state. The analysis recovers the expected result that one can indeed associate the free energy difference $\Delta F$ to an arbitrary pure qubit state $\ket{\psi}\bra{\psi}$, but only within a particular ``classical regime'', in which one has access to an infinite system with unbounded coherence resources. Outside of this setting it is provably impossible to extract all of the free energy from the quantum coherence.

The actual implementation of these thermodynamic processes often assume a complex protocol. A great deal of control is required over the different components in order to magnify the energy acquisition up to scales in which the notion of ordered, robust energy makes more sense. The central aim of this paper to address scenarios in which such thermodynamic processes are carried out on a quantum system $S$ via a quasi-autonomous thermal machine $M$ that is \emph{comparable in scale} and itself displays quantum-mechanical properties. This sheds light on the physical characteristics demanded of a quantum thermal machine. 

In the case of work extraction, a finite-sized quantum machine will absorb energy which, due to its finite dimension, will diminish its ability to function. We show that a resolution to this is to perform an ``energy harvesting measurement'' on the machine. This harvesting measurement serves a dual function of siphoning off energy from the machine while also stabilizing its ability to function as a thermal machine at these quantum-mechanical scales.

\subsection{Clocks and quantum coherence}
As highlighted in \cite{Lostaglio2014,Oppenheim2014aa,Lostaglio2015,Korzekwa}, in order for thermodynamic processes to be sensitive to quantum coherence at arbitrary scales the thermal machine must itself possess coherent properties. In \cite{Lostaglio2015} this requirement for quantum coherence was identified with transformations that break time-translation symmetry. Any quantification of the coherence follows from a quantification of this asymmetry. 

More explicitly, the breaking of time-translation symmetry demands that the effects of an action depend on whether it is performed at time $t_1$ or at some later time $t_2 > t_1$. The way in which one handles time-translation asymmetry has been known for a long time -- we introduce a ``clock'' system, sensitive to the passage of time \cite{0517377446}. In the classical regime we of course have abundant access to time-keeping devices, however this becomes a more non-trivial component for extremely small, autonomous quantum devices, or in environments in which quantum-mechanical aspects dominate.

Such considerations have also resulted in an increased focus on the role of clocks in thermodynamics \cite{Salmilehto2014,Malabarba2014a,Rankovi,Frenzel2014d,Korzekwa,Series2015,Tajima2015}. To implement autonomous quantum thermodynamic protocols, generic thermal machines invariably require a clock degree of freedom, which serves as a non-classical time-keeping device. This not only allows access to quantum coherence, but also can be used to induce effective time-dependent interactions within systems.

\subsection{Energy transfer to macroscopic scales via quantum measurements}
The use of the clock in, say, a work-extraction protocol necessitates a non-trivial interaction between the quantum system $S$ and the clock. The unavoidable back-action experienced by the machine is in general accompanied by an energy flow that can either be transferred to other degrees of freedom in the machine $M$, or more simply maintained in the clock itself. However if the machine $M$ is comparable in scale to the quantum system $S$ then it is debatable to what extent one has ``gained work'' if it is confined to quantum-mechanical degrees of freedom in $M$. A basic requirement is that the acquired energy can be transferred to larger scales in some natural manner.

The passage from the quantum regime to the classical regime has long been a topic of controversy and debate. Where does quantum end and classical begin? Notions that are applicable on the classical side of this divide are inapplicable on the quantum side, and quantum measurements play a central role in linking these two regimes. Depending on which side of the cut one places the measurement device one can either view a measurement as an abrupt transformation of the quantum state (e.g.\!\! via a projective measurement), or one could equally model the measurement process itself as a purely unitary interaction between the quantum system $S$ and some measurement apparatus $A$. The purpose of the unitary interaction $U_{SA}$ is to magnify the quantum-mechanical aspects up the ladder of length-scales to degrees of freedom that are deemed classical. The apparatus eventually admits a read-off and an objective measurement outcome is obtained.

One particular set of constraints when considering fully quantum mechanical measurement approaches is given by conservation laws. The connection between conservation laws and measurements has a long and subtle history. The WAY-theorem states that in the presence of a conservation law there is an effective superselection rule in place on the observables that can be measured \cite{Wigner1952,Araki1960,Yanase1961,Ahmadi2013,Marvian2012,Navascues2014,Loveridge2011}. In \cite{Ahmadi2013} it was shown that the measurement device must carry two different resources -- a coherence resource to partially lift the superselection rule, and a charge degree of freedom to balance books. The unitary interaction between the machine $M$ and the system $S$ is constantly entangling the two systems. This unitary \emph{correlation process} can be viewed as an \emph{information-acquisition} by the machine in an effective local energy basis that varies with time. Note that this process of $M$ acquiring information on $S$ is distinct from the \emph{external} measurement performed on $M$ we will later consider.

In what follows, we view the operation of the thermal machine $M$ with the system $S$ and reservoir $R$ as transforming an energetic degree of freedom of the device under $\rho_S\otimes \rho_M \otimes \gamma_R \rightarrow U_{SM} (\rho_S \otimes \rho_M\otimes \gamma_R) U_{SM}^\dagger$, which can then read-off via a subsequent measurement on $M$. This disturbing measurement transfers the energy acquired by the quantum thermal machine into the macroscopic regime, where it can be ascribed a less ambiguous status.

In what follows we use thermodynamic work extraction from a qubit system as our focus. The work extraction involves a quasi-autonomous thermal machine based on a \emph{globally time-independent} Hamiltonian, which performs the protocol on the quantum side of the cut, together with a continual flow of energy across the cut via a classically-controlled measurement process that also serves to stabilize the quantum device.

\section{Autonomous Quantum Machines:\\ The Basic Constituents}

\begin{figure}
\includegraphics[width=6cm]{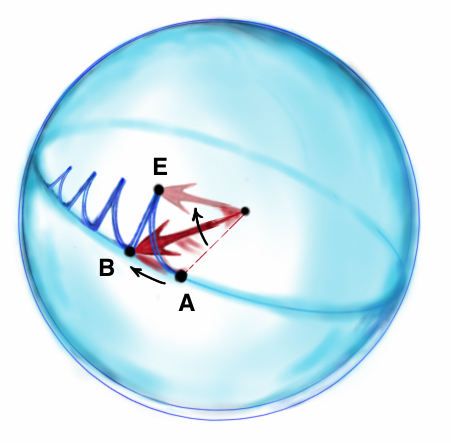}
\caption{\textbf{Dynamics of the thermal machine.} The quantum thermal machine at point $\boldsymbol{A}$ of its orbit, absorbs energy from the heat reservoir, which causes a fraction of the total state to deviate from the reference clock orbit. Specifically a fraction $p_1(t)$ evolves to point $\boldsymbol{E}$ off the reference orbit, while the remaining fraction $(1-p_1(t))$ freely evolves to $\boldsymbol{B}$. The energy-harvesting measurement projects the component at $\boldsymbol{E}$ onto $\boldsymbol{B}$ and allows us to extract this energy as well as stabilizing the machine back to the reference orbit. Each triangular section constitutes a single unit protocol (UP). The total entropy generated corresponds to the integrated ``area'' under the jagged curve, while the extracted work corresponds to the integrated ``length'' of the jagged curve. The reference orbit $\chi(t)$ is the shaded great circle with respect to which the protocol is defined.\label{fig:clock}}
\end{figure}

The traditional Szilard argument begins with the knowledge that the system is in one particular state of a pair of energetically degenerate levels (e.g.\! a particle is on the left-hand side of a piston) \cite{Szilard1929,Kim2011,Plenio2001,Alicki2004,Frenzel2014d,Renner2011}. Conditioned on this knowledge, the agent applies an adapted protocol that extracts a certain quantity of work. Schematically the unoccupied energy level is elevated by switching on a time-dependent Hamiltonian (e.g.\! attaching weights, or tuning magnetic fields). The system is then placed in equilibrium with a thermal reservoir at some inverse temperature $\beta$, and the Hamiltonian is quasi-statically switched off in a classically controlled manner.

In the fully quantum-mechanical scenario our starting point is again knowing that the qubit system $S$ is in a definite pure state $\ket{\psi}\bra{\psi}$. We next feed this state into a thermal machine $M$, and the composite evolves under the joint Hamiltonian $H_{SM}$, together with system-bath couplings. Crucially, note that we do \emph{not} assume classical control over the interaction between the quantum systems. Therefore, to obtain a non-trivial extraction of energy requires us to \emph{induce} a particular time-dependent evolution of the system $S$ -- therefore the machine $M$ must possess a clock degree of freedom that induces an \emph{effective} time-varying Hamiltonian on the target system. In \cite{Rankovi} the authors also consider quantum systems that via interactions with their environment act as clocks, and study their synchronisation as well as the back-action they suffer due to the interaction, although not in a thermodynamic setting. The clock in \cite{Rankovi} comprises two components, a \emph{clockwork} which evolves due to an internal mechanism, and \emph{tick registers} which briefly interact with the clockwork and extract time information. The notion of a clockwork in \cite{Rankovi} is similar to what we call the clock itself, while the system $S$ can loosely be seen as playing the role of the tick registers. The clock in the present study however is in continuous contact with the system and evolves with it under a joint unitary.

\subsection{Induced local time-dependent level splittings}
The known state $\ket{\psi}\bra{\psi}$ defines an orthonormal basis $\{\ket{\psi}, \ket{\overline{\psi}} \}$ in the qubit system. Given this preferred decomposition, the generic interaction Hamiltonian needed on the joint Hilbert space $\H_{SM}$ takes the form
\begin{eqnarray}\label{eq:HJoint}
H_{SM} = \sigma \otimes H_{I}  + \I \otimes H_{F}
\end{eqnarray}
where $\sigma :=  \ket{\overline{\psi}}\bra{\overline{\psi}} - \ket{\psi}\bra{\psi}$ (see appendix \ref{app:Details}). In particular, the term $\sigma \otimes H_{I}$ is what will generate a level-splitting local to the qubit, with the time-dependence being encoded in the thermal machine's quantum state $\rho_M$. The second term $\I \otimes H_F$, as we shall see, generates an evolution on the system $M$ which is not sensitive to the state of the qubit, and so can be interpreted as the free Hamiltonian of the machine.
Crucially, this joint Hamiltonian is \emph{time-independent}, fixed for all eternity.

In the absence of bath couplings, the system-machine composite evolves under this Hamiltonian as $\rho_{SM} \rightarrow e^{-i t H_{SM} } \rho_{SM} e^{i t H_{SM}}$.
The function of the interaction with the machine is to induce an effective Hamiltonian that is local to the qubit. We define
\begin{equation}\label{eq:HReducedDef}
H_S(t) := \mbox{tr} _M[ \I \otimes \rho_M(t) H_{SM}],
\end{equation}
which is a time-dependent mean-field Hamiltonian on the qubit. This mean-field approximation turns out to be the physically appropriate choice in the context considered, and has been analysed in more detail in \cite{Frenzel2014d,Teifel2011}. For the above interaction Hamiltonian this takes the form
\begin{eqnarray}\label{eq:HReduced}
H_S(t) = \braket{H_{I}} \sigma + \braket{H_{F}}\mathbb{1}.
\end{eqnarray}
This mean-field Hamiltonian defines a \emph{local effective basis}, and encodes the non-correlative (and therefore non-entropy increasing) dynamics local to the system \cite{Weimer2007,Schroder2010,Teifel2011}. Crucially, the thermal bath is assumed to only see this local Hamiltonian, which encodes the statement that the machine $M$ does not undergo direct thermalization. This assumption is equivalent to demanding a large coupling strength between system and bath relative to the coupling to the machine, such that the thermalisation time-scale of the system is much shorter than that of the machine. The exact degree to which thermalization of $S$ occurs will depend on the particular bath coupling rates (see also appendix \ref{app:therm}).

Since the energy exchanges with the bath only depend on the level-splitting of $H_S$ it suffices to assume $H_S(t) = \braket{H_{I}} \sigma$. Also, note that instead of time being an explicit parameter in the system Hamiltonian tuned by an external agent, the time-dependence is now induced by the dynamics and the particular quantum state of the machine $M$, giving the machine an inbuilt quantum clock. The time-dependence is explicitly a function of the coherence properties of the state $\rho_M$ with respect to the Hamiltonian $H_{SM}$, as emphasized previously\footnote{We use the terms clock and machine interchangeably, for the discussion here the precise distinctions between the clock degree of freedom and the rest of the machine is not important.}.

Therefore, the joint Hamiltonian is essentially determined for the (known) quantum state $\ket{\psi}\bra{\psi}$ and, together with an initial joint state
\begin{eqnarray}\label{eq:rhoJointInitial}
\rho(0) =  \ket{\psi}\bra{\psi}\otimes \rho_M(0),
\end{eqnarray}
induces a specific time-dependent level-splitting local to the qubit. However one must also address the back-action of the dynamics on the machine itself. The joint Hamiltonian can be written in the alternative form
\begin{eqnarray}\label{eq:HJointSplit}
H_{SM} = \ket{\psi}\bra{\psi} \otimes H_{-}+ \ket{\overline{\psi}}\bra{\overline{\psi}} \otimes H_{+}
\end{eqnarray}
where we have defined $H_{\pm} := H_{F} \pm H_{I}$. The evolution that is generated by this Hamiltonian splits into two parts
\begin{eqnarray}\label{eq:UTotal}
U(t) = \ket{\psi}\bra{\psi}  \otimes U_-(t) + \ket{\overline{\psi}}\bra{\overline{\psi}} \otimes U_+ (t)
\end{eqnarray}
where $U_{\pm}(t) = e^{-iH_{\pm} t}$. This describes a controlled unitary action, in which conditioning on the qubit's $\ket{\psi}$ and $\ket{\overline{\psi}}$ states evolves the clock along two independent orbits according to $U_-$ and $U_+$ respectively.\\

\subsection{Designing a good machine: core requirements}
We now turn to the question of what clock characteristics besides the validity of eq. (\ref{eq:HReducedDef}) are desirable for the functioning of the thermal machine. To make this a well-posed problem, we demand that the clock's Hilbert space dimension is fixed and that the Hamiltonian has its spectrum upper bounded by some fixed energy-scale $||H_{SM} || \le E$, but we are otherwise free in designing the machine's Hamiltonians and its starting state $\rho_M(0)$.

Firstly, the Szilard argument requires an initial degeneracy in the energies of the qubit. Secondly, it is desirable (but not essential) to fix the energy of the state $\ket{\psi}$ to be zero, so that the qubit's induced level-splitting according to eq. (\ref{eq:HReduced}) is given by $\Delta(t) := \mbox{tr}[\rho_M(t)H_{+}]$. Finally and most importantly we require the right coherence properties of the machine to ensure that it functions well, both as a clock and in its ability to induce level-splittings on the qubit.
These three criteria are respectively encoded by the following set of conditions on the operators $\{ \rho_M(0), H_-, H_+\}$:
\begin{Lalign}
\mbox{tr}  [ \rho_M(0)  H_I ] &= 0, \tag{i}\label{cond1} \\
 \mbox{tr}  [ \rho_M(0)  H_- ] & = 0, \tag{ii}\label{cond2} \\
 \mbox{Im} \left (\mbox{tr} \bigl[\rho_M(0)[H_-, H_+] \bigr] \right )& \gg 0 .  \tag{iii}\label{cond3}
\end{Lalign} 
Conditions (\ref{cond1}) and (\ref{cond2}) follow directly and uniquely from the desire for initial degeneracy and fixing of the ground state. The intuition behind (\ref{cond3}) is less straight forward and deserves some elaboration. We would like the Hamiltonians $H_+$ and $H_-$ to have a large commutator in an operator norm sense. At the level of the algebraic relations this implies a rapidly changing unitary evolution (as can be seen from expanding the unitary in increasing orders of commutators). But we then require the state to have a strong response to the induced dynamics. This can be achieved by having the initial state ``mutually unbiased'' with respect to the Hamiltonians. Note that every Hilbert space $\mathcal{H}$ admits a triple of mutually unbiased bases. An extreme regime is the case of $H_{\pm}$ being built from two of these bases, while the quantum state is one of the basis states in the third basis. This guarantees that the state simultaneously has maximal coherence with respect to both bases and therefore will strongly break time-translation invariance. Note that taking the imaginary part in (\ref{cond3}) is due to the fact that the trace will be imaginary due to the anti-Hermitian nature of the commutator. While not being unique, condition (\ref{cond3}) provides a convenient encapsulation of these physical requirements.

From eq. (\ref{eq:HJointSplit}) we see that in the absence of any thermal contact, the qubit remains in the state $\ket{\psi}$ for all time as the joint system freely evolves under $H_{SM}$. This defines a reference trajectory for the clock, whose dynamics are fully determined by $U_-$. We therefore define $\chi(t):=U_-(t)\rho_M(0) U_-^{\dagger}(t)$ as the clock state at time $t$ on the ideal \emph{reference clock-orbit}. Clearly, for any given Hamiltonian $H_{SM}$, there is a range of states $\{\rho_M^{(m)}(0)\}_m$ that satisfy conditions (\ref{cond1}) and (\ref{cond2}) and ensure validity of eq. (\ref{eq:HReducedDef}). Each of these initial clock states has its own unique clock-orbit $\chi^{(m)}(t)$ associated with it. We choose the specific state $\rho_M(0)$ as the state whose associated orbit $\chi(t)$ maximises the level splitting $\Delta(t)$ for some $t=\tilde{\tau}$ over the orbit, i.e. 
\begin{equation}\label{eq:ChiT}
\chi(\tilde{\tau}) := \argmax_{\chi^{(m)}} \left ( \max_t \mbox{tr}[\chi^{(m)}(t)H_+] \right).
\end{equation}
This maximising state can always be taken to be a pure state which we call $\ket{d}$, where $d$ refers to the Hilbert space dimension of the clock. The gap maximization then requires that $U_-$ rotates this pure state $\ket{d}\bra{d}$ (which obeys conditions (\ref{cond1}) and (\ref{cond2})) into the maximum eigenvalue eigenstate of $H_+$. To achieve this, it is sufficient to design the machine's Hamiltonians such that $H_+$ and $H_-$ are generators of SU(2) on the $d$-dimensional machine (see appendix \ref{app:SU2} for details). Assuming this choice, eq. (\ref{eq:ChiT}) is equivalent to optimising for condition (\ref{cond3}), so that the optimal machine starting state is the maximum eigenvalue eigenstate of the operator $C:=i[H_-, H_+]$ (which is also an SU(2) generator). We define the eigenbasis of $C$ with ascending eigenvalues as $\{\ket{m}\}_{1 \leq m \leq d}$, such that $\rho_M(0) = \ket{d}\bra{d}$. Finally, we introduce the complete rotating clock basis $\{\ket{m(t)} := U_-(t)\ket{m}\}$ which co-rotates with the clock's $m=d$ reference orbit $\chi(t) = \ket{d(t)}\bra{d(t)}$. 

Choosing $H_-$ to be an SU(2) generator  on the system Hilbert space $\H$ also has the advantage of ensuring closed, periodic orbits and will allow us to run the engine in a well-defined cycle\footnote{If we do not choose such an $H_-$, we have to impose an additional condition, demanding the ratio of the eigenvalues of $H_-$ to be rational. Note that any real number can be approximated by a rational number with arbitrary accuracy, and thus this restriction could be considered as unnecessarily strict since any clock will always be periodic to arbitrary accuracy. However, its period might approach infinity in these cases, so the restriction makes sense from a practical view point.}. We call the period of the clock $\tau$, such that $\tau$ is the smallest positive number for which $U_-(\tau) = \mathbb{1}$. The $d$ different orbits swept out by the $\{\ket{m(t)}\}$ clock basis states over one clock period form a zero-energy surface with respect to the qubit's $\ket{\psi}$ state. Another important property is that, for all $t=n\tau$ with $n\in \mathbb{N}$, the orthogonal states $\ket{m(n\tau)}$ have equal energy with respect to \emph{any} qubit state. This follows since the entire eigenbasis of $C$ satisfies condition (\ref{cond1}), and so we may switch between clock orbits without any energy cost at these times. 

\section{The Explicit Protocol}
Having established the basic quantum-mechanical properties required of the engine, we now turn to the details of the actual engine protocol. First there is the initialization phase in which the initial state (\ref{eq:rhoJointInitial}) freely evolves for a time $\tilde{\tau}$ at which point it attains a maximal local level-splitting $\Delta_{max} := \Delta(\tilde{\tau})$. This induces a local raising of the system's unoccupied $\ket{\overline{\psi}}$ state, as in the original Szilard-type protocol.  However the interaction Hamiltonian $H_{SM}$ is fundamentally time-independent, and so there is a need for a single timing-switch within the machine that initialises thermal contact between the qubit and the thermal bath at time $t= \tilde{\tau}$. It is not entirely clear that such a single time switching is fundamentally necessary, but in approaches such as the present one it appears to be almost impossible to avoid if one wants to keep the model general.

With thermal contact in place, work extraction then takes place for $\tilde{\tau} < t \leq \tau$ and can be analysed in steps of duration $dt$, which can be viewed as constituting \textit{unit protocols} (UP). Each UP can be further analysed in terms of three sub-components: 
\begin{enumerate}[(a)]
\item Thermalisation of the system $S$. \label{step1}
\item Global dynamical evolution of system and machine. \label{step2}
\item Harvesting measurement on the machine $M$. \label{step3}
\end{enumerate}
It is important to note that the division into these units is determined by the \emph{macroscopically-controlled} timing of measurements, and not at the level of the quantum machine.

\subsection{Autonomy of the Thermal Machine}

One might argue that the frequent measurements in a (as shall be seen later) time-dependent basis in step (\ref{step3}) are in fact similar to a non-autonomous machine having a fine-tuned time-dependent Hamiltonian which is externally controlled. Note though that these two scenarios only become comparable in the limit $dt\rightarrow 0$ where the measurements occur at a very high rate. However, our model is valid for arbitrary $dt$, and one can even consider the extreme case in which $dt=\tau-\tilde{\tau}$ and only a single measurement in a fixed basis is performed at the end of the protocol\footnote{In this case one has to consider more realistic thermalisation protocols such as the ones discussed in appendix \ref{app:therm}. The choice of a single thermalisation per $dt$, as employed in the main text, is only necessary for obtaining the analytical expressions, but not a fundamental property of the model.}. Despite a reduced work output and higher probability of failure (see below), the thermal machine is still able to extract work from the system, even though the process in this limit becomes comparable to the standard two-point work measurement \cite{PhysRevE.75.050102,RevModPhys.83.771} employed in many quantum thermodynamic protocols, but with a fixed Hamiltonian. In the non-autonomous scenario if the Hamiltonian was fixed no work output would be possible in this case. Since the machine studied here is able to extract work for any numbers of interventions we call it \emph{quasi-autonomous}. It abstractly coincides with the fully non-autonomous case only in the limit of continuous intervention $dt\rightarrow 0$.

\subsection{Dynamics of System \& Thermal Machine}
The exact thermalization process can be modelled in various ways, including non-trivial interaction with the unitary dynamics generated by $H_{SM}$. However, for the sake of analysis we may approximate steps (\ref{step1}) and (\ref{step2}) as firstly a thermalization of the qubit with respect to the local mean-field Hamiltonian $H_S(t)$, followed by unitary dynamics $\exp [ -i H_{SM}d t ]$. This approximation is robust over a large range of parameters, and exact in the $dt \rightarrow 0$ limit (see appendix \ref{app:therm} and discussion in \cite{Frenzel2014d}).

As such, at the beginning of every UP the joint system is to good approximation in a state $\rho(t) = \gamma(t)\otimes\chi(t)$, where $\gamma(t)=p_0(t)\ket{\psi}\bra{\psi} + p_1(t)\ket{\overline{\psi}}\bra{\overline{\psi}}$ is the qubit's Gibbs state with respect to its local, mean-field Hamiltonian $H_S(t)$ (eq. (\ref{eq:HReduced})) at inverse temperature $\beta$, such that $p_0(t) = (1+e^{-\beta \Delta(t)})^{-1}$ and $p_1(t) = 1-p_0(t)$.

The evolution under the Hamiltonian $H_{SM}$ takes the joint system from $\rho(t)$ to $\rho'(t+dt) := U(dt)\rho(t)U^{\dagger}(dt)$. However, since the qubit is generically in a mixed state, the clock now \emph{deviates} from the reference clock orbit\footnote{As an illustrative analogy, one can think of the state $\chi(t)$ as the hand of a clock and $dt$ as the fundamental unit of time, a 'second'. If the qubit has a non-zero $\ket{\overline{\psi}}$ component, instead of simply ticking to the next position $\chi(t+dt)$, the hand of the clock splits up into two parts ending up in a convex mixture of the expected $\chi(t+dt)$ state, and the state $U_+(dt)\chi(t)U_+^{\dagger}(dt)$.}, i.e. $\mbox{tr}_S[\rho'(t+dt)] \neq \chi(t+dt)$, as schematically illustrated in Fig. \ref{fig:clock}. Crucially, this deviation of the clock from its reference orbit corresponds to energy being transferred from the qubit system into the machine. Since the state of the clock is distorted by this gain it therefore not only acts as a time-keeping device but also as \emph{temporary} battery.

\subsection{Energy-harvesting and clock stabilization}

Since the quantum thermal machine suffers a back-action as it absorbs energy its ability to induce local-level splittings and to function as a clock is affected. A crucial component of the protocol is that we repeatedly perform energy-harvesting measurements on the machine that serve two functions: firstly to transfer the energy gain from the quantum to the classical regime, and secondly to stabilize the clock/machine system. This in turn allows us to separate the concepts of clock and battery in the quantum-mechanical system.

The target state on the reference clock orbit is given by $\chi (t+dt)$, and therefore the measurement we perform is the projective rank-1 measurement in the orthonormal clock basis $\{\ket{m(t+dt)}\}$. The ability to perform this measurement is assumed to be a free operation that is accessible macroscopically, however this too could be modelled more explicitly using a larger coherent reference, if one wished. When the measurement is performed, with high probability we project back onto the reference orbit $\chi(t+dt)$, thus stabilizing the clock. This probability tends to one as we either decrease the thermal couplings or increase the rate $dt^{-1}$ at which we perform the measurements (see appendices \ref{app:Details} and \ref{app:Zeno}). 

The measurement performed during each UP does not commute with $H_{SM}$ and is thus not energy-conserving. It therefore leads to energy flows between the joint system and the external measurement device. Since energy is globally conserved we can explicitly compute the energy flow into the measurement device. For the outcome corresponding to the projection $\Pi_m (t+dt) := \I \otimes \ket{m(t+dt)} \bra{m(t+dt)} $ we find this to be 
\begin{eqnarray} \label{eq:dEGeneral}
dE_m(t,dt) = \mbox{tr}\Bigl[H_{SM}\bigl(\rho'(t+dt) - \rho_m(t+dt)\bigr) \Bigr],
\end{eqnarray}
where $\rho_m (t+dt) \propto \Pi_m(t+dt)  \rho'(t+dt)\Pi_m(t+dt) $ is the post-measurement state of the system and machine conditioned on outcome $m$.

\subsection{Exorcising Demons:\\Landauer accounting in the macroscopic regime}

While eq. (\ref{eq:dEGeneral}) provides the exchange of energy between the quantum device and measurement apparatus, we 
cannot identify this as the extracted work. The reason is that the measurement device must be re-set and its memory erased in order to avoid any Maxwell's demon type scenarios \cite{maxwell1872theory, Earman1998,Maruyama2009,Zurek1990,Chapman2015,Shiraishi2015,Bub2001,Mandal17072012,PhysRevLett.111.030602}. However this memory is a classical record in the macroscopic apparatus and therefore the traditional Landauer cost of erasure applies \cite{Landauer1961,Bennett2003,Hilt2011,Faist2012,Goold2014}. 

The information gain by the apparatus is described by the distribution of measurement outcomes $\p(t):= \{p_m (t,dt)\}$, and so the minimal cost of erasure is given by
\begin{eqnarray} \label{eq:dWResetGeneral}
dW_{reset}(t,dt) = \beta^{-1} S\bigl( \p (t) \bigr),
\end{eqnarray}
where $S(\cdot )$ is the Shannon entropy. We can therefore identify an averaged work gain during the UP as 
\begin{eqnarray} \label{eq:dWGeneral}
\hspace{-0.6cm}dW(t,dt) = \sum_m p_m(t,dt) dE_m(t,dt) - dW_{reset}(t,dt).
\end{eqnarray}

This provides a basic energy accounting over the unit protocol from time $t$ to time $t+dt$. However in order to compose multiple UPs over the entire engine cycle requires us to address the matter of what happens when we do not obtain the projection onto the reference clock orbit, corresponding to $m =d$.

\subsection{Modes of operation and quantum feedback}

We now restrict our focus to two distinct operation modes for the engine, which we refer to as the \textit{unselective} and \textit{selective} protocols.

In the unselective case a sequence of measurements is performed in intervals of length $dt$, and the outcomes recorded. Finally at the end the total measurement data is erased and we have a net energy gain which is the work output. For this we can define the trajectory $\mathbf{m} := \{m_{\tilde{\tau}}, m_{\tilde{\tau}+dt},...,m_{\tau}\}$ as the set of measurement outcomes of the $N=\frac{\tau-\tilde{\tau}}{dt}$ consecutive UPs of a full cycle. In each UP, the energy flow into the measurement device given measurement outcome $m_{t+dt}$, conditioned on starting the UP in state labelled by $m_t$ is
\begin{eqnarray}\label{eq:dEConditional}
\!\!\!\!\! dE(m_{t+dt} | m_t) = \mbox{tr}\Bigl[H_{SM}\Delta\rho(m_{t+dt} | m_{t})\Bigr]\hspace{-0.05cm},
\end{eqnarray}
where $\Delta\rho(m_{t+dt} | m_{t}) := \rho(m_{t})-\rho(m_{t+dt} | m_{t})$ is the difference between the initial state of the UP and the post-measurement state. The notation $\rho(m_{t})$ shows explicitly that the clock does not necessarily start on the reference orbit, but on any of the orbits labelled by $1\leq m_t \leq d$, and that the qubit is in a thermal state with respect to the local Hamiltonian induced by this orbit. It is important to note that the energy flow only depends on the clock state directly preceding each UP, not the entire trajectory, since the thermalisation step essentially kills any trajectory history resulting in a Markov process (see appendix \ref{app:Details}). 

The reset at the end of of the cycle is given by the Landauer expression. If $p(\mathbf{m})$ denotes the probability of a certain complete trajectory $\mathbf{m}$, then the reset cost is given by $W_{reset} = \beta^{-1}S(\{p(\mathbf{m})\})$. 
The average work output of the unselective engine is
\begin{eqnarray}\label{eq:dWTrajectoryAverage}
\braket{W}_u = \sum_{\mathbf{m}}p(\mathbf{m}) E(\mathbf{m}) - \beta^{-1}S(\{p(\mathbf{m})\}),
\end{eqnarray}
where $E(\mathbf{m}) = \sum_{n=1}^N dE(m_{\tilde{\tau}+ndt} | m_{\tilde{\tau}+(n-1)dt})$ is the energy flow for the trajectory $\mathbf{m}$. The unselective protocol constitutes a minimalist approach, in which the quantum components require no feedback control. It is therefore the most autonomous mode of operation.

However in this unselective regime the thermal machine undergoes non-trivial back-action that degrades the clock. One might therefore wish to allow elementary feedback control on the quantum systems with the aim of maintaining the characteristics of the machine. Feedback control has been extensively studied in the context of work extraction protocols, both in the classical as well as in the quantum case (see e.g. \cite{Funo2013,Sagawa2008,Ito2013,Horowitz2011,Horowitz2014,Ashida2014}). These feedback protocols generally employ measurements of the target system, followed by operations which are chosen based on the specific result of the measurement. Similarly here, in the selective protocol we operate conditional on the measurement outcomes. If $m=d$ we have a successful projection onto the reference clock orbit, and all is well. The clock is restored back into the state $\chi(t+dt)$, successfully stabilising it, and the joint state is $\rho_d(t+dt) =  \rho_S' (t+ dt) \otimes \chi(t+dt)$. 

For any $m\neq d$ a non-ideal outcome occurs and the clock jumps into a different orbit $\ket{m(t+dt)}$. Moreover, if one reads off the outcome, then the qubit is collapsed into $\ket{\overline{\psi}}$, resulting in a joint state $\rho_m(t+dt) = \ket{\overline{\psi}}\bra{\overline{\psi}} \otimes\ket{m(t+dt)}\bra{m(t+dt)}$ (see appendix \ref{app:Details} for details and the specific form of $\rho_S'(t+dt)$). The quantum engine has ``misfired'', and in this case we abort the current engine cycle, decouple the qubit from the bath, and perform the following feedback process that resets the engine for a new cycle. The feedback on the system $S$ flips  $\ket{\overline{\psi}}$ into $\ket{\psi}$ through a single qubit unitary (whose energy cost has to be accounted for), followed by the free evolution of the joint system, ``running the engine in neutral'', for a duration $\tau-(t+dt)$ so that the clock ends up in $\ket{m(\tau)} = \ket{m(0)}$. Crucially, since for $t=n\pi$ all states of the clock basis have equal energy as noted above, we may now restore the clock to the reference orbit $\chi(0)$ without any energy cost, ready to begin a new cycle. 

\section{Actual Performance}

The cumulative erasure cost in the selective mode is generally smaller than the single large erasure in the unselective mode. Starting from the work expression in eq. (\ref{eq:dWGeneral}) one can compute the average work output of this engine mode similar to eq. (\ref{eq:dWTrajectoryAverage}). This average, which we shall simply call $\braket{W}$, needs to take into account the probabilities of the engine succeeding each UP, as well as the cost of the feedback protocol in case the engine misfires. The explicit expression for $\braket{W}$ eq. (\ref{WAverageGeneral}) is derived in appendix \ref{app:energy}. We can also define $W_{ideal}$ as the maximum single-shot work output of a cycle that completes without a misfire. Although not established explicitly for general clocks, the selective engine employing a feedback protocol has a higher work output $W_{ideal} \ge \braket{W}\ge\braket{W}_u$ in all examples considered, with equality in the Zeno limit, which we shall now consider.

\subsection{Thermodynamic reversibility and the Zeno Limit}

The limiting case of $dt\rightarrow 0$ constitutes a Zeno limit (ZL) and takes on a special role, since it allows us to recover the well-established results of equilibrium thermodynamics. Explicitly evaluating all the quantities involved, it is easy to show (see appendix \ref{app:Zeno}) that the probability of being projected back into the reference clock orbit is equal to unity up to first order in $dt$. Thus the selective engine will complete the entire cycle without a single UP failing with probability $1-\mathcal{O}(dt^2)$, and the unselective engine will follow the reference orbit trajectory $\{d,d,...,d\}$ with a probability $1-\mathcal{O}(dt^2)$. It can also be shown that in both cases the cost of resetting vanishes up to first order, $W_{reset} = \mathcal{O}(dt^2)$, making the process essentially reversible.  This implies that in the ZL both engine modes are equivalent. These results further imply that eq. (\ref{eq:dWGeneral}) reduces to $\lim_{dt\rightarrow 0} dW(t,dt) = dE(t,dt)$, i.e. the entire energy flowing into the measurement device can actually be identified as work.\\
More specifically, we find for the infinitesimal work flow
\begin{equation}\label{eq:dWExplicit}
dW(t,dt) = -i p_1(t) \mbox{tr}\Bigl[\chi(t) [H_- , H_+]\Bigr] dt +  \mathcal{O}(dt^2),
\end{equation}
where $p_1(t)$ is the qubit's thermal occupation with respect to the local Hamiltonian induced by $\chi(t)$. 

We can compare this to the change in free energy of the qubit. Its partition function is $Z(t) = 1 + e^{-\beta\Delta(t)}$. Substituting this into the infinitesimal change in free energy $dF(t) = d(-\beta\ln Z(t))$ we recover the quasi-static equilibrium result,
\begin{eqnarray} \label{eq:dEGeneralZeno}
\lim_{dt\rightarrow 0} dW(t,dt) = - dF(t).
\end{eqnarray}
Specifically, the quantum Zeno engine with a built-in clock, constantly stabilised via energy-harvesting measurements, is able to extract the entire free energy difference of the system as work. Conversely, if we are not able to perfectly stabilise the clock at all times and only allow the accumulated energy to flow out of the quantum clock and into the classical battery at finite intervals, we are naturally restricted to $\Delta W < -\Delta F$. This is in accord with the second law, where equality can only be achieved under reversible protocols. 

This also demonstrates a core tradeoff between work output and power. On the assumption that we are experimentally restricted to a minimum $dt$, we can either attempt to slow down the system dynamics to get closer to the ZL at the expense of power, or vice versa get a higher power but being further from equality in eq. (\ref{eq:dEGeneralZeno}), ``wasting'' free energy. From eq. (\ref{eq:dWExplicit}) we can also see that if the qubit's $\ket{\overline{\psi}}$ state is lowered too fast compared to $dt$, the clock is not able to sample the qubit's thermal distribution $p_1(t)$ quick enough to utilize the full free energy difference. This adds to the tradeoff between maximising power and maximising work output.

Integrating eq. (\ref{eq:dEGeneralZeno}) over $\tilde{\tau}\leq t \leq \tau$, the total work output of the Zeno engine is
\begin{eqnarray} \label{eq:WGeneralZeno}
W_{Zeno} &=& kT \Bigl(\log2 - \log Z(\tilde{\tau}) \Bigr),
\end{eqnarray}
where $Z(\tilde{\tau})$ is the partition function of the qubit at maximum level splitting $\Delta(\tilde{\tau})$. This shows a second limitation we suffer if considering a realistic finite-sized machine. Even if we were able to perfectly stabilise the clock, we are further limited by the maximum level-splitting that the clock can induce. Only in the limit of an infinitely big (i.e. classical) clock can we reach an infinite level splitting $\Delta(\tilde{\tau})\rightarrow\infty$ (i.e. $Z(\tilde{\tau})\rightarrow1$) and thus obtain the classical result $W = kT\log2$. 

\begin{figure}[tb]
\begin{center}
\includegraphics[width = \columnwidth]{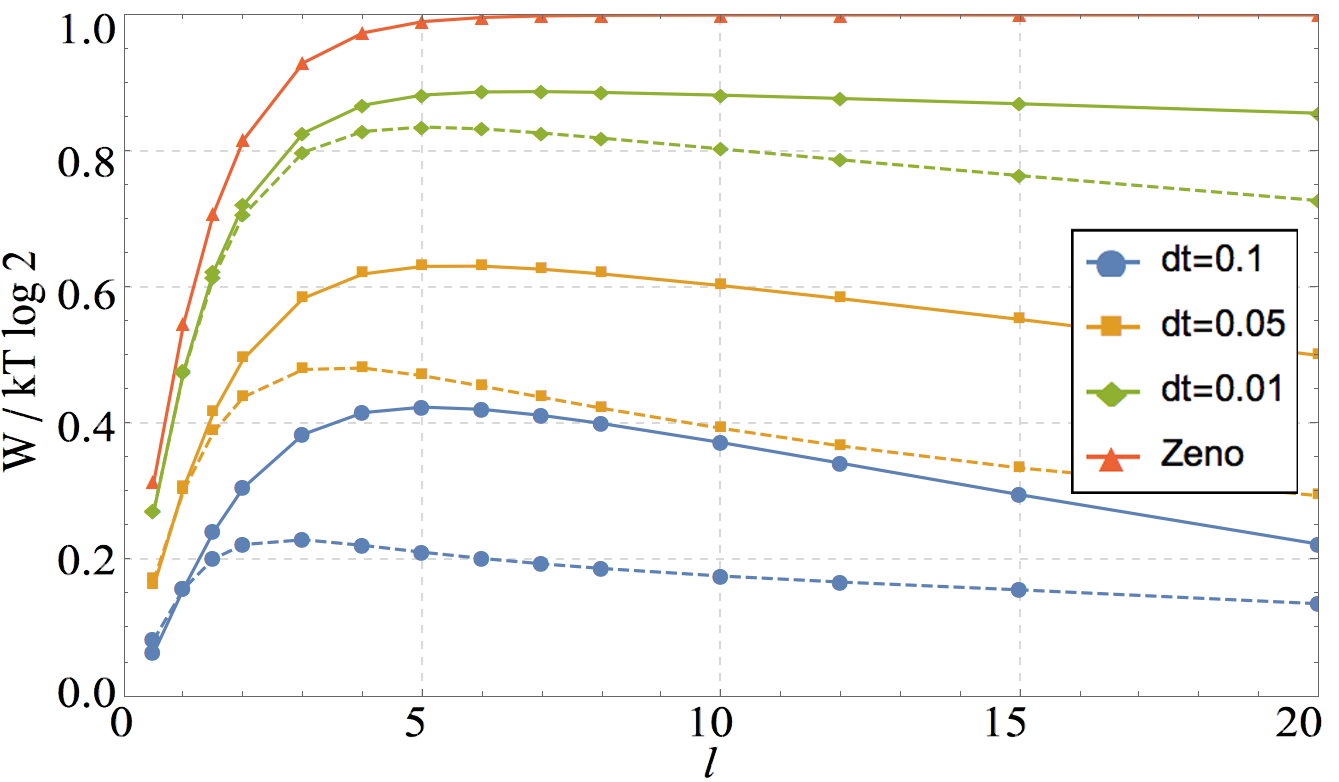}
\caption{Ideal $W_{ideal}$ (solid) and average $\braket{W}$ (dashed) work output for a spin-clock-driven engine against clock size $l$ for different stabilisation intervals $dt$. The red curve shows the Zeno limit $dt\rightarrow0$ which quickly approaches the classical result $W=kT\log2$.}
\label{fig:SpinWork}
\end{center}
\end{figure}
\section{An Example: The Spin Clock}
The preceding discussion has remained abstract, and not tied to a specific physical realisation. It provides a broad framework of quasi-autonomous quantum engines driven by measurement-stabilised clocks. However, we can look more closely at an explicit example where a spin-$l$ particle (dimension $d=2l+1$) acts as the clock \cite{Frenzel2014d}. The joint-Hamiltonian takes the form $H_{SM}=\frac{1}{\sqrt{2}}(\sigma\otimes L_z + \mathbb{1}\otimes L_y)$ which implies $H_{\pm} = (L_y \pm L_z)/\sqrt{2}$, where $L_k$ is the angular momentum operator of the spin-$l$ particle along the $k$-axis. We notice that as desired $H_-$ and $H_+$ are also generators of SU(2) and find the third generator $C=i [H_- , H_+] = -L_x$. As shown in the general framework, the optimal initial clock state $\chi(0)$ is the eigenstate of $C$ with maximum eigenvalue, i.e. a spin fully polarised along the negative $x$-direction. The full clock-basis comprises the eigenbasis of $-L_x$. 

This example is particularly nice since the spin can be viewed as the quantum-analogue of a clock hand. Under free evolution the state $\chi(t)$ simply rotates around an axis defined by $H_-$, just like a clock hand with period $2\pi$. The other clock-orbits co-rotate with the reference orbit, and can essentially be seen as shortened, fuzzy versions of the clock hand, i.e. the spin not being fully polarised in a certain direction but only partially. Note that if there is no polarisation, for example if the clock gets too mixed, the hand disappears, we are unable to tell the time, and so unable to induce time-dependence in the qubit. The effect of backaction from the qubit on the clock is again a stochastic splitting of the clock hand into two parts, one following the clock-orbit, the other one rotating out of the clock-plane.

Applying the earlier analysis to this specific model, one can explicitly calculate the work output of a spin-clock-driven quantum engine for varying spin dimensions $d$ and stabilisation intervals $dt$.
\begin{figure}[tb]
\begin{center}
\includegraphics[width = \columnwidth]{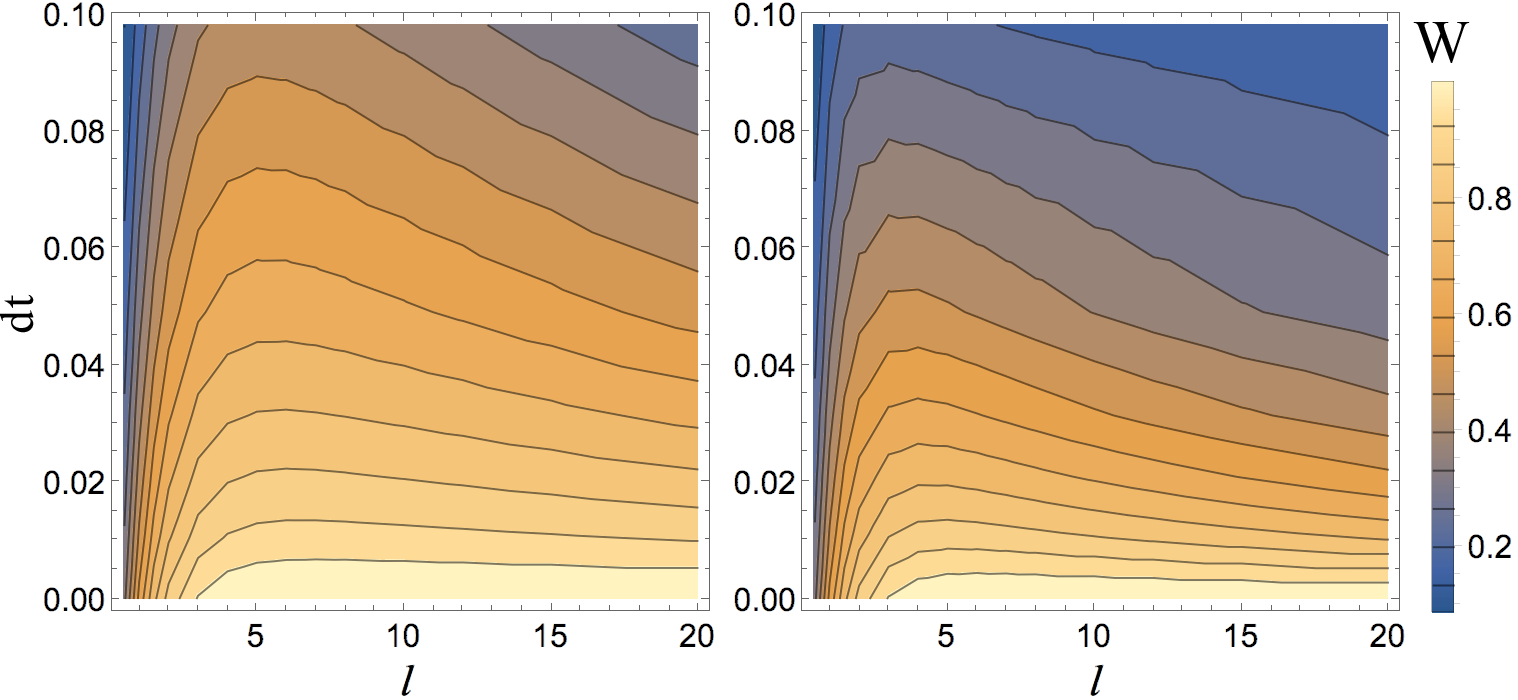}
\caption{Contours of work output (normalised to $kT\log2$) of the selective engine for clock sizes $l$ and stabilisation interval $dt$. The left plot shows the ideal scenario in which every measurement succeeds. The right plot is the average work output of the selective engine.}
\label{fig:WorkContour}
\end{center}
\end{figure}
The results are shown in Figures \ref{fig:SpinWork} and \ref{fig:WorkContour}. We see that in the ZL even for small clocks we quickly saturate the classical result $W=kT\log2$. What might be surprising at first is that for finite $dt$ there is an optimal $d$ which is also finite. This comes back to the issue of sampling the qubit. If the clock is too small, we are unable to raise the $\ket{\overline{\psi}}$ state high enough before starting to thermalise. On the other hand, if the clock is too large, we raise the state very high, but also drop it very fast and hence might not be able to sample the thermal distribution quick enough. For larger clocks the average quickly falls below the ideal output, but then converges again in the $l\rightarrow\infty$ limit since in this case the optimal scenario starts to dominate the average due to the increasingly smaller deviation from the reference orbit experienced by larger clocks.\\
Again we see that only in the limit of an infinitely large clock with infinitely fast stabilisation can the classical limit of $W = kT\log2$ be recovered.

\section{Outlook}

We have presented an explicit analysis of the basic requirements of a quasi-autonomous quantum thermal machine\footnote{Note that while our discussions were focused on a machine operating on pure qubits, the same derivations generalise to arbitrary $d$-dimensional systems and mixed input states (appendix \ref{app:mixed}).}. Its sole purpose is to extract work from a qubit degree of freedom. 

The operation of the thermal machine involves a complex interplay between certain core properties such as quantum coherence, clock degrees of freedom, and the role of quantum measurements. It would be of theoretical interest to provide more detailed accounts of these aspects as they relate to thermodynamics. While we have a good framework to understand the stochastic aspects of thermodynamics \cite{Brandao2013}, we do not have the same level of clarity regarding the role that coherence has on thermodynamic processes. Such a problem can be tackled from a variety of directions, including the present approach.

Beyond this, it would be valuable to make greater connection with current experimental progress related to this line of inquiry. Already there has been work in the context of optomechanical systems \cite{Elouard2013,Arcizet2011,Kippenberg2008,Yeo2014}, and our work has direct bearing on these models. 

From the theoretical perspective it would also be fruitful to further develop the information feedback components, and to consider quantum memory degrees of freedom at the level of the quantum thermal machine. In particular it would be valuable to explore the role that measurements play in such systems and to provide a fuller account for generic POVM scenarios. Ultimately this provides a useful setting to address the interplay between information and energy flows across the quantum-classical divide.

\section{Acknowledgments}
We acknowledge useful discussions with Kais Abdelkhalek and David Reeb. We also thank Janet Anders for helpful feedback on an earlier draft. This work was partially supported by the COST Action MP1209. MF supported by the EPSRC and the Japan Society for the Promotion of Science, and DJ supported by the Royal Society.

\bibliography{QZE.bib}

\appendix

\section{Designing a machine with SU(2) Hamiltonians}\label{app:SU2}
In the main text it was stated that choosing $H_-$ and $H_+$ to be SU(2) generators, finding the clock orbit via eq. (\ref{eq:ChiT}) becomes equivalent to finding the initial state which optimises the coherence condition (\ref{cond3}). Using the definition $C:=i[H_-,H_+]$ we can rewrite condition (\ref{cond3}) as
\begin{eqnarray} \label{eq:cond1}
\mbox{tr} [\rho_M(0)C] \gg 0.
\end{eqnarray}
This condition is optimised by choosing $\rho_M(0) =\ket{d}\bra{d}$ where $\ket{d}$ is the maximum eigenvalue eigenstate of $C$. Now, since $H_-$ and $H_+$ are SU(2) generators, the definition of $C$ implies that it is also a generator of SU(2), and one can think of $H_-$, $H_+$ and $C$ as representing three orthogonal axes. Since the reference clock-orbit evolution is given by $\chi(t)=\exp[-iH_- t]\ket{d}\bra{d}\exp[iH_- t]$, i.e. rotating a state that lies initially along the axis defined by $C$ around the orthogonal axis defined by $H_-$, we are guaranteed that for some time $t=\tilde{\tau}$, the state $\chi(\tilde{\tau})$ will coincide with the eigenstate of $H_+$ which optimises eq. (\ref{eq:ChiT}).

Due to the symmetry properties of SU(2) (the generators essentially defining orthogonal axes), condition (\ref{cond2}) is satisfied for any state $\ket{m}$ in the eigenbasis of $C$, not only the optimal state $\ket{d}$, since the expectation values of $H_+$ and $H_-$ with respect to the eigenstates of $C$ are zero. Thus, since the clock orbits are defined with evolution under $U_-$ only, and $U_-$ and $H_-$ commute, condition (\ref{cond2}) is in fact satisfied at all times if the clock is on any of the reference orbits $\ket{m(t)}$. Thus the set of $d$ clock orbits $\{\ket{m(t)}\}_{1\leq m \leq d}$ defines a zero energy surface with respect to the qubit's $\ket{\psi}$ state.

Finally, since as noted above $\braket{m | H_- | m} = 0 = \braket{m | H_+ | m}$ for all $m$, and $H_I = \frac{1}{2}(H_+ - H_-)$, condition (\ref{cond1}) is also satisfied for all eigenstates of $C$. This condition does however in general only hold at the initial time $t=0$ and after a complete period $t=\tau$. The fact that for times $0<t<\tau$ this condition is broken allows us to induce time-dependent level-splittings with respect to the clock orbits, while the fact that it holds for any complete period $t=n\tau$ allows us to freely switch between the different $\ket{m}$ states and reset the engine to the reference orbit without any energy cost in case it was projected on a different orbit during the cycle.

These considerations show that, while not necessary, it is very desirable to design the quantum thermal machine $M$ in such a way that $H_\pm$ are SU(2) generators.

\section{The protocols in detail} \label{app:Details}
In this section we shall analyse the engine protocol outlined in the main text in more detail. The protocol starts with the joint system-machine system in the state $\rho(0) = \ket{\psi}\bra{\psi}\otimes\chi(0)$.

The initialisation phase consists of free evolution of the initial state $\rho(0)$ for a time $\tilde{\tau}$ at which point the qubit's induced level-splitting 
\begin{equation}
\Delta(t) = \mbox{tr}[\rho_M(t)H_+]
\end{equation}
attains a maximum $\Delta_{max} = \Delta(\tilde{\tau})$. This free evolution leaves the qubit state unchanged and rotates the clock state into $\chi(\tilde{\tau})=\ket{d(\tilde{\tau})}\bra{d(\tilde{\tau})}$, the maximum eigenvalue eigenstate of $H_+$. 

We now start the actual work extraction process which takes place from $\tilde{\tau}\leq t \leq \tau$ and is broken down into $N=\frac{\tau - \tilde{\tau}}{dt}$ subroutines of duration $dt$, which we call unit protocols (UP). These subroutines are further broken down into three steps as outlined in (\ref{step1})-(\ref{step3}).

\subsection{Thermalisation}

The first step of each UP is thermalisation of the qubit with respect to its local Hamiltonian $H_S(t)$ induced by the state of the clock. Assuming that the clock is on the reference orbit at time $t$, the joint state after thermalisation is 
\begin{equation}
\rho(t) = \gamma(t) \otimes \chi(t),
\end{equation}
where $\gamma(t)=p_0(t)\ket{\psi}\bra{\psi} + p_1(t)\ket{\overline{\psi}}\bra{\overline{\psi}}$ is the Gibbs state with respect to the reference clock orbit at time $t$ such that $p_0(t) = (1+e^{-\beta \Delta(t)})^{-1}$ and $p_1(t) = 1-p_0(t)$ and $\Delta(t)=\braket{d(t)|H_+|d(t)}$.

If on the other hand the previous measurement projected the clock onto a different orbit $m$ and we keep the engine running (as is the case in the unselective protocol), the qubit's local mean-field Hamiltonian takes on a different form, giving it a different level-splitting and thus thermalising to a different Gibbs state $\gamma(t | m)=p^{(m)}_0(t)\ket{\psi}\bra{\psi} + p^{(m)}_1(t)\ket{\overline{\psi}}\bra{\overline{\psi}}$, where the thermal probabilities are defined as above but with respect to the level-splitting $\Delta(t|m)=\braket{m(t)|H_+|m(t)}$, leading to a post-thermalisation joint state of
\begin{equation} \label{eq:rhoTM}
\rho(t | m) = \gamma(t|m) \otimes \ket{m(t)}\bra{m(t)}.
\end{equation}
Note that $\rho(t|d)=\rho(t)$, but we make the distinction to keep notation in the case of the selective engine more concise.

\subsection{Evolution}

Step (\ref{step2}) of the UP consists of free evolution of the joint state for a duration $dt$. Note that for brevity we will in the following assume that all evolution operators act for a time $dt$, such that e.g. $U_- \equiv U_-(dt)$, unless stated otherwise. Under this evolution the state $\rho(t)$ evolves into $\rho'(t+dt) = U\rho(t)U^{\dagger}$. More explicitly,
\begin{eqnarray}
\rho'(t+dt) =& p_0(t)\ket{\psi}\bra{\psi}\otimes U_-\chi(t) U_-^{\dagger} \nonumber \\
&+ p_1(t)\ket{\overline{\psi}}\bra{\overline{\psi}}\otimes U_+\chi(t) U_+^{\dagger}.
\end{eqnarray}
The first term corresponds to the clock simply evolving along the reference clock orbit for a time $dt$, since $U_-\chi(t) U_-^{\dagger} = \chi(t+dt)$. The second term however corresponds to a deviation of the clock from the reference orbit, and an injection of energy into the clock. The expression for $\rho'(t+dt|m) = U\rho(t|m)U^{\dagger}$ has a similar form and interpretation, only with the reference clock orbit $\chi(t)$ replaced by the co-rotating orbit $\ket{m(t)}\bra{m(t)}$.\\

\subsection{Measurement in the selective mode}

Finally in the last step (\ref{step3}) of the UP we perform a measurement to try and stabilise the clock and transfer the energy it acquired during step (\ref{step2}) to the macroscopic measurement apparatus. The measurement that will project the clock back onto one of the clock orbits is described by the projectors $\Pi_m (t+dt) := \I \otimes \ket{m(t+dt)} \bra{m(t+dt)}$. Applying this to the state $\rho'(t+dt)$, the post measurement state given measurement outcome $m$ is 
\begin{widetext}
\begin{equation}
\rho_m(t+dt) = 
\begin{cases} \label{eq:rhoMExplicit}
\frac{1}{p_d(t+dt)} \left (p_0(t) \ket{\psi}\bra{\psi} + p_1(t)\Gamma_{dd}(t,dt) \ket{\overline{\psi}}\bra{\overline{\psi}} \right) \otimes \chi(t+dt) &\mbox{ if } m=d \\
\ket{\overline{\psi}}\bra{\overline{\psi}}\otimes \ket{m(t+dt)}\bra{m(t+dt)}  &\mbox{ if } m\neq d,
\end{cases}
\end{equation}
with probability
\begin{equation} \label{eq:pMsel}
p_m(t,dt) = 
\begin{cases} 
p_0(t) + p_1(t)\Gamma_{dd}(t,dt) &\mbox{ if } m=d \\
p_1(t)\Gamma_{md}(t,dt) &\mbox{ if } m\neq d,
\end{cases}
\end{equation}
\end{widetext}
where we have defined
\begin{equation} \label{eq:Gamma}
\Gamma_{m'm}(t,dt) := \left| \braket{ m'(t+dt) | U_+ | m(t)} \right|^2.
\end{equation}
This quantity $\Gamma_{m'm}(t,dt)$ can be understood as the probability of the clock transitioning into the orbit labelled by $m'$ under the deviation-inducing evolution $U_+$ for duration $dt$, given that the the clock was in the orbit labelled by $m$ at time $t$. It can easily be verified that the $\Gamma_{m'm}$ form a doubly-stochastic matrix, with summation over either of the two indices giving unity. Also note that if $dt$ is small, this matrix is diagonally dominant, i.e. the system is more likely to remain on any given orbit than to transition to a different orbit.

From eq. (\ref{eq:rhoMExplicit}) we can directly see that if we observe the measurement outcome $m=d$, the clock has been successfully stabilised and projected back onto the reference clock orbit at $\chi(t+dt)$. The qubit is steered to a slightly altered state $\rho'_S(t+dt) \propto p_0(t) \ket{\psi}\bra{\psi} + p_1(t)\Gamma_{dd}(t,dt) \ket{\overline{\psi}}\bra{\overline{\psi}}$. On the other hand, if we observe $m\neq d$, the clock transitions to a different orbit (often leading to a backflow of energy from the measurement apparatus as we shall show below) and the qubit is instantly collapsed to $\ket{\overline{\psi}}$.

In the selective operation mode of the engine, such a ``misfire'' of the engine, a measurement outcome $m\neq d$, triggers an abortion of the current engine cycle and a feedback procedure that resets the engine for a new cycle. The qubit is immediately decouple from the bath to avoid further thermalisation. Further, we need to reset the qubit to the $\ket{\psi}$ state to ensure that the clock rotates along the clock-orbit and ends up in a state where it can be restored to the reference orbit without any further energy cost. The flip operation is given by the unitary $U_{f} := (\ket{\psi}\bra{\overline{\psi}} + \ket{\overline{\psi}}\bra{\psi}) \otimes \mathbb{1}$ and takes the state $\rho_m(t+dt)$ to $\rho'_m(t+dt) = \ket{{\psi}}\bra{{\psi}}\otimes \ket{m(t+dt)}\bra{m(t+dt)}$. This flip is not energy conserving and its cost has to be taken into account (see below). Since the qubit is now in the $\ket{\psi}$ state again, we can allow the system to freely evolve for a duration $\tau-(t+dt)$, resulting in a state $ \ket{{\psi}}\bra{{\psi}}\otimes \ket{m(\tau)}\bra{m(\tau)}$. Now, due to condition (\ref{cond1}), we can restore the clock to the reference orbit at $\chi(0)$ for free, and the joint system is ready to begin a new cycle.

\subsection{Measurement in the unselective mode}
In the case of the unselective operation mode the state after measurement depends both on the measurement outcome $m'$ at time $t+dt$, as well as the the previous measurement outcome $m$ at time $t$. Explicitly, the post-measurement state after observing $m'$ is given by
\begin{widetext}
\begin{equation}
\rho(m' | m) = 
\begin{cases} \label{eq:rhoMExplicit}
\frac{1}{p(m' | m)} \left (p_0^{(m)}(t) \ket{\psi}\bra{\psi} + p_1^{(m)}(t)\Gamma_{m'm}(t,dt) \ket{\overline{\psi}}\bra{\overline{\psi}} \right) \otimes \ket{m'(t+dt)}\bra{m'(t+dt)} &\mbox{ if } m'=m \\
\ket{\overline{\psi}}\bra{\overline{\psi}}\otimes \ket{m'(t+dt)}\bra{m'(t+dt)}  &\mbox{ if } m'\neq m,
\end{cases}
\end{equation}
with probability
\begin{equation} \label{eq:pMun}
p(m'|m) = 
\begin{cases} 
p_0^{(m)}(t) + p^{(m)}_1(t)\Gamma_{mm}(t,dt) &\mbox{ if } m'=m \\
p_1^{(m)}(t)\Gamma_{m'm}(t,dt) &\mbox{ if } m'\neq m.
\end{cases}
\end{equation}
\end{widetext}
If the engine is run in the unselective mode, the engine is kept running regardless of the specific outcome $m'$, and the measurement is followed by the next UP, beginning with a new thermalisation to the state $\rho(t+dt | m')$ eq. (\ref{eq:rhoTM}), which destroys the information of the previous measurement, resulting in a Markov process.

\section{Energy flows} \label{app:energy}
Knowing all the quantum states of the system $S$ and machine $M$ during the entire protocol it is straight forward to calculate the energy flows into the measurement apparatus and the thermal reservoir by invoking global energy conservation.

\subsection{Work}

The energy flow into the measurement apparatus takes place during the measurement process. It is given by the energy difference of the pre- and post-measurement quantum states. In the selective mode, given measurement outcome $m$ we have for the energy flow into the macroscopic apparatus
\begin{eqnarray}
dE_m(t,dt) = \mbox{tr}\Bigl[H_{SM}\bigl(\rho'(t+dt) - \rho_m(t+dt)\bigr) \Bigr].
\end{eqnarray}
Substituting the explicit expressions for the states and invoking condition (\ref{cond2}) we find for the energy flow in the case of the ideal measurement outcome $m=d$
\begin{eqnarray}\label{eq:dEd}
dE_d(t,dt) = p_1(t) \braket{ d(t) | \Delta H_+(t,dt) | d(t)},
\end{eqnarray}
where we have defined
\begin{equation}\label{eq:DeltaHP}
\Delta H_+(t,dt) := H_+ - \frac{\Gamma_{dd}(t,dt)}{p_d(t,dt)} U_-^{\dagger}H_+U_-.
\end{equation}
Since $0 < \Gamma_{dd}(t,dt) < 1$, the ratio 
\begin{equation}\label{eq:GammaRatio}
\frac{\Gamma_{dd}(t,dt)}{p_d(t,dt)} = \frac{1}{p_1(t) + p_0(t)/\Gamma_{dd}(t,dt)} < 1
\end{equation}
is strictly less than $1$. In addition the expectation value of $H_+$ is greater than the expectation value of $U_-^{\dagger}H_+U_-$ in eq. (\ref{eq:dEd}), leading to a positive energy flow into the apparatus\footnote{Note that this is not strictly true for every UP in general if the clock evolves along a complicated trajectory (i.e. if the level splitting $\Delta(t)$ is not monotonically decreasing), but is always true on average over the interval $\tilde{\tau}<t<\tau$.}.

However, in the case of a misfire event, a measurement outcome $m\neq d$, we have a very different energy flow, specifically
\begin{eqnarray}\label{eq:dEm}
dE'_{m\neq d}(t,dt) =& &p_1(t) \braket{ d(t) | H_+ | d(t)} \nonumber \\&-& \braket{ m(t+dt) | H_+ | m(t+dt)}.
\end{eqnarray}
The dash indicates that this is only part of the energy flow associated with this event (see below). Even though the expectation value of $H_+$ with respect to the state $\ket{m(t+dt)}$ is generally less than that with respect to $\ket{d(t)}$, the fact that $0<p_1(t)\leq\frac{1}{2}$ implies that in many cases an engine misfire implies a back-flow of energy from the apparatus into the quantum system\footnote{Note that again $0<p_1(t)\leq\frac{1}{2}$ is not necessarily true in general for complex clock evolution since $\Delta(t)$ can in principle get negative, but it holds on average over the interval $\tilde{\tau}<t<\tau$.}. Additionally applying the feedback and flipping the qubit via the unitary $U_f$ will lead to an additional energy cost (which we assume is taken from the work stored in the apparatus)
\begin{eqnarray}\label{eq:dEflip}
dE_{m}^{f}(t,dt) = -\braket{ m(t+dt) | H_+ | m(t+dt)}.
\end{eqnarray}
Taking this feedback cost into account, the total energy exchange between quantum system and measurement apparatus given a measurement outcome $m\neq d$ is 
\begin{eqnarray}\label{eq:dETot}
dE_{m\neq d}(t,dt) =& &p_1(t) \braket{ d(t) | H_+ | d(t)} \nonumber \\&-& 2\braket{ m(t+dt) | H_+ | m(t+dt)}.
\end{eqnarray}
Also taking into account the cost of resetting the memory $W_{reset}(t,dt) = \beta^{-1} S\bigl( \p (t) \bigr)$ given in eq. (\ref{eq:dWResetGeneral}), we arrive at the average work output of the UP starting at time $t$ in the selective mode of
\begin{eqnarray} 
\hspace{-0.5cm}dW(t,dt) = \sum_m p_m(t,dt) dE_m(t,dt) - dW_{reset}(t,dt).
\end{eqnarray}
We can split this up and define the work associated with outcome $m$ as
\begin{eqnarray} \label{eq:dWm}
\hspace{-0.5cm}dW_m(t,dt) := dE_m(t,dt) - dW_{reset}(t,dt),
\end{eqnarray}
such that $dW = \sum_m p_m dW_m$. Using this, we find for the total work output of the selective engine averaged over a full engine cycle
\begin{widetext}
\begin{eqnarray}\label{WAverageGeneral}
\braket{W} = &&\Bigl(\prod_{n=1}^N p_d(\tilde{\tau}+n dt)\Bigr)\sum_{n=1}^N dW_d(\tilde{\tau}+n dt) \nonumber \\
&+& \sum_{k=2}^{N-1}\Bigl[ \prod_{n=1}^{k-1} p_d(\tilde{\tau}+n dt)\Bigr](1-p_d(\tilde{\tau}+k dt)) \sum_{n=1}^{k-1} dW_d(\tilde{\tau}+n dt)\nonumber\\
&+& \sum_{k=1}^N\Bigl[ \prod_{n=1}^{k-1} p_d(\tilde{\tau}+n dt)\Bigr] \sum_{m=1}^{d-1}p_m(\tilde{\tau}+k dt)dW_m(\tilde{\tau}+k dt),
\end{eqnarray}
\end{widetext}
where the first term corresponds to the work extracted in the case each of the $N$ unit protocols succeeding, the second term corresponds to the work extracted up to a misfire at the $k$-th UP (which also aborts the cycle), and the final term contains the energy flow of the misfire at the $k$-th UP itself, all weighted by the respective probabilities of these events occurring.

For the unselective engine mode, we find the energy associated with with a transition from the clock orbit $m$ at time $t$ to the orbit $m'$ at time $t+dt$ (c.f. eq. (\ref{eq:dEConditional})) as
\begin{eqnarray}\label{eq:dEConditionalExplicit}
\!\!\!\!\! dE(m' | m) = p_1^{m}(t)& \bigl(& \braket{m|H_+|m} \nonumber \\ &-&\frac{\Gamma_{m'm}}{p(m'|m)}\braket{m'|H_+|m'}\bigr),
\end{eqnarray}
where we have omitted the explicit time dependence for notational brevity.

The probability of a complete engine trajectory $\mathbf{m}=\{m_{\tilde{\tau}}, m_{\tilde{\tau}+dt},...,m_{\tau}\}$ over all $N$ unit protocols is
\begin{equation}
p(\mathbf{m}) = p(m_{\tilde{\tau}})\prod_{n = 1}^N p(m_{\tilde{\tau}+ndt} | m_{\tilde{\tau}+(n-1)dt}),
\end{equation}
where the inclusion of $m_{\tilde{\tau}}$ in the trajectory allows for the scenario in which the clock begins the protocol on a orbit other than the reference orbit. The total energy flow over this trajectory is 
\begin{equation}
E(\mathbf{m}) = \sum_{n=1}^N dE(m_{\tilde{\tau}+ndt} | m_{\tilde{\tau}+(n-1)dt}).
\end{equation}
Finally at the end of the engine cycle we have to reset the memory which is associated with a work cost $W_{reset} = \beta^{-1}S(\{p(\mathbf{m})\})$, such that the total average work output of the unselective engine is given by
\begin{eqnarray}\label{eq:dWTrajectoryAverageApp}
\braket{W}_u = \sum_{\mathbf{m}}p(\mathbf{m}) E(\mathbf{m}) - \beta^{-1}S(\{p(\mathbf{m})\}).
\end{eqnarray}
where we have averaged over all possible engine trajectories.

\subsection{Heat}
During step (\ref{step1}) of each UP the qubit thermalises with respect to its induced local Hamiltonian by interacting with the thermal reservoir at inverse temperature $\beta$. Before the thermalisation at time $t+dt$ the qubit is in the state 
\begin{eqnarray}
\rho'_S(t+dt) &=& \frac{p_0(t)}{p_d(t,dt)} \ket{\psi}\bra{\psi} + \frac{p_1(t)\Gamma_{dd}(t,dt)}{p_d(t,dt)} \ket{\overline{\psi}}\bra{\overline{\psi}} \nonumber \\
&:=& p_0^* \ket{\psi}\bra{\psi} + p_1^* \ket{\overline{\psi}}\bra{\overline{\psi}}
\end{eqnarray}
as can be seen from eq. (\ref{eq:rhoMExplicit}). Since in general $p_0(t+dt) < p_0(t)$ and $p_1(t+dt) > p_1(t)$ (since $\Delta(t+dt)<\Delta(t)$), and $p_0^* > p_0(t)$ and $p_1^* < p_1(t)$, we see that the interaction of the qubit with the machine, and the back-action of the measurement process on the machine actually drive the qubit even further away from its thermal state at time $t+dt$ than it would have been otherwise, leading to an increased heat flow required to thermalise the qubit. Specifically the heat flow during the thermalisation at time $t+dt$ is given by
\begin{eqnarray}
dQ(t+dt) &=& \mbox{tr}\Bigl[H_{SM}\bigl(\rho(t+dt) - \rho_d(t+dt)\bigr) \Bigr]  \\
&=& \left (p_1(t+dt) - p_1(t)\frac{\Gamma_{dd}(t,dt)}{p_d(t,dt)}\right) \Delta(t+dt).\nonumber
\end{eqnarray}
The heat flow in the classical case is given by the same expression but without the ratio $\Gamma_{dd}(t,dt)/p_d(t,dt) < 1$, showing that the fully-quantum mechanical protocol has a higher heat flow associated with it. However, as we hall show below, this ratio approaches $1$ in the Zeno limit, such the the classical result can be recovered even for finite-size quantum machines, as long as they can be stabilised infinitely fast. A similar result can be derived for the unselective mode of operation.

\section{Zeno Limit derivations} \label{app:Zeno}
In this section we derive the results for the Zeno limit $dt\rightarrow0$, in which the clock is stabilised with infinite fidelity, and energy flows constantly out of the quantum machine into the measurement apparatus. 

Let us first consider the quantity $\Gamma_{m'm}$ defined in eq. (\ref{eq:Gamma}). We can rewrite it as
\begin{equation}
\Gamma_{m'm}(t,dt) = \left|\braket{m'(t) | e^{iH_- dt}e^{-iH_+ dt} | m(t)}\right|^2.
\end{equation}
Expanding the exponential functions ignoring terms of order $\mathcal{O}(dt^2)$ or higher we have
\begin{eqnarray} \label{eq:Gammadt}
\Gamma_{m'm}(t,dt) &\approx& \left|\braket{m'(t) | (\mathbb{1} + iH_- dt)(\mathbb{1} - iH_+ dt) | m(t)}\right|^2 \nonumber \\
&\approx& \left|\delta_{m'm} + i\braket{m'(t) | H_- - H_+ | m(t)}dt\right|^2 \nonumber \\
&=& \delta_{m'm} - \mathcal{O}(dt^2).
\end{eqnarray}
Thus the probability of the clock staying on a certain orbit $m$, despite the deviating inducing evolution generated by $H_+$, is equal to unity up to first order in $dt$. Crucially this includes the clock orbit $d$ such that $\Gamma_{dd}(t,dt) = 1 - \mathcal{O}(dt^2)$.

Having established the limiting value of this central quantity, we can now consider the probability distribution over measurement outcomes $\mathbf{p}(t)$. Specifically, we are interested in the probability associated with the clock being projected back onto it's reference orbit, assuming it started the UP in this orbit, i.e. the probability $p_d(t,dt)$ eq. (\ref{eq:pMsel}) in the selective case, and $p(d|d)$ eq. (\ref{eq:pMun}) in the unselective case. We first note that $p_d(t,dt) = p(d|d)$. We have
\begin{eqnarray}\label{eq:pddt}
p_d(t,dt) &=& p_0(t) + \Gamma_{dd}(t,dt)p_1(t) \nonumber \\
&=& p_0(t) + p_1(t) - \mathcal{O}(dt^2) \nonumber\\
&=& 1 - \mathcal{O}(dt^2)
\end{eqnarray}
Thus in the Zeno limit we are guaranteed to project back onto the reference orbit and the selective and unselective protocols become equivalent. From this result it also immediately follows that the cost of resetting the memory vanishes for both operation modes as $W_{reset} = \mathcal{O}(dt^2)$. Thus from eq. (\ref{eq:dWGeneral}) we have 
\begin{equation}\label{eq:dWdt}
dW(t,dt) = dE_d(t,dt) - \mathcal{O}(dt^2).
\end{equation}
To find an expression for $dE_d(t,dt)$ in the $dt\rightarrow0$ limit we see from eqs. (\ref{eq:dEd}) and (\ref{eq:DeltaHP}) that we need to evaluate the expression $\frac{\Gamma_{dd}(t,dt)}{p_d(t,dt)} U_-^{\dagger}H_+U_-$. For the ratio we find $\frac{\Gamma_{dd}(t,dt)}{p_d(t,dt)} = 1-\mathcal{O}(dt^2)$ as can be verified by substituting eqs. (\ref{eq:Gammadt}) and (\ref{eq:pddt}). For the other factor we have upon expanding the exponential functions
\begin{eqnarray}
U_-^{\dagger}H_+U_- &=& (\mathbb{1} + iH_- dt)H_+(\mathbb{1} - iH_- dt) + \mathcal{O}(dt^2)\nonumber \\
&=& H_+ - iH_+H_- dt + i H_- H_+ dt + \mathcal{O}(dt^2) \nonumber\\
&=& H_+ + i[H_-,H_+]dt + \mathcal{O}(dt^2)
\end{eqnarray}
from which it follows that
\begin{equation}
\Delta H_+(t,dt) = -i[H_-,H_+]dt + \mathcal{O}(dt^2).
\end{equation}
Finally, substituting this expression into eq. (\ref{eq:dEd}) and the result into eq. (\ref{eq:dWdt}) we arrive at the result for $dW(t,dt)$ eq. (\ref{eq:dWExplicit}).

Let us now focus on the free energy change of the qubit to prove eq. (\ref{eq:dEGeneralZeno}), showing that in the Zeno limit the entire free energy difference can be extracted as work. Given the fact that we fixed the qubit's $\ket{\psi}$ state to zero energy via condition (\ref{cond2}), the partition function at time $t$ is given by $Z(t) = 1 + e^{-\beta \Delta(t)}$ where $\Delta(t) = \mbox{tr}[\chi(t) H_+]$, where we have assumed that the clock is on the clock orbit $\chi(t)$ as we showed is always the case in the Zeno limit. The change in free energy of the qubit is thus
\begin{eqnarray}\label{eq:dF}
dF(t) &=& d(-\beta^{-1} \log Z(t))\nonumber \\
&=& -\beta^{-1}  \frac{1}{Z(t)}\frac{dZ}{dt} dt.
\end{eqnarray}
Differentiating the partition function with respect to $t$ we get
\begin{eqnarray}
\frac{dZ}{dt} &=& -\beta \frac{d\Delta}{dt} e^{-\beta \Delta(t)}.
\end{eqnarray}
The energy splitting of the qubit varies in time as
\begin{eqnarray}
\frac{d\Delta}{dt} &=& \frac{d}{dt}\mbox{tr}[\chi(t) H_+] \nonumber\\
&=& \frac{d}{dt}\mbox{tr}[U_-(t)\chi(0)U_-^{\dagger}(t) H_+] \nonumber\\
&=& i\mbox{tr}\left[\chi(t) [H_-,H_+]\right].
\end{eqnarray}
Finally, recognising that $\frac{e^{-\beta \Delta(t)}}{Z(t)} = p_1(t)$ and substituting everything into eq. (\ref{eq:dF}), we see that $dF(t) = -dW(t,dt)$ up to first order in $dt$, which concludes the proof of eq. (\ref{eq:dEGeneralZeno}).

Integrating eq. (\ref{eq:dF}) over the interval $\tilde{\tau} \leq t \leq \tau$ and noting that $Z(\tau) = 2$ due to the qubit's degeneracy at $t=\tau$, eq. (\ref{eq:WGeneralZeno}) follows immediately.

\section{Detailed analysis of the spin-clock} \label{app:Example}
As outlined in the main text, for the specific example of a spin-$l$ system acting as quantum machine/clock we choose the Hamiltonian
\begin{equation}
H_{SM}=\frac{1}{\sqrt{2}}(\sigma\otimes L_z + \mathbb{1}\otimes L_y),
\end{equation}
such that $H_{\pm} = \frac{1}{\sqrt{2}}(L_y \pm L_z)$, where $L_k$ is the angular momentum operator of the spin-$l$ particle along the $k$-axis. The angular momentum operators $L_k$ clearly are SU(2) generators, and the operators $H_{\pm}$ can also be seen as angular momentum operators defining a new coordinate system. The third SU(2) generator can be found via the commutation relation 
\begin{equation}
C=i [H_- , H_+] = -L_x.
\end{equation}
Thus we see that this new coordinate system has essentially been flipped along the $x$-axis as well as rotated in the $y$-$z$-plane. The ideal clock state we want to choose is the maximum eigenvalue eigenstate of $C$, i.e. a spin fully polarised along the negative $x$-direction. We define the eigenbasis $\{\ket{m}\}_{-l\leq m \leq l}$ such that $C\ket{m} = m\ket{m}$. Note that here we use a slightly different convention from the remainder of the text where $1\leq m \leq d$. This is more suited to the angular momentum eigenbases. The state on the reference clock orbit at time $t$ is thus given by
\begin{equation}
\chi(t) = U_-(t) \ket{l}\bra{l} U_-^{\dagger}(t).
\end{equation}
Considering the rotation generated by $U_-$ (or using the more formal Wigner matrix derivation introduced below), it is straightforward to show that the level splitting of the qubit induced by a clock on this orbit is simply
\begin{equation}
\Delta(t) = l\sin t.
\end{equation}
We see that the maximum level splitting that this clock can induce is $\Delta_{max}=l$ at a time $\tilde{\tau}=\pi/2$. Note also that even though the period of the clock is technically $\tau=2\pi$, it is preferable to stop the machine earlier at $\tau' = \pi$, since the qubit is degenerate again at this time with respect to all clock orbits, and keeping the engine running for the remaining period would at best (namely in the Zeno limit) lead to no additional work gain. Even though this was not explicitly stated in the main text, whenever the qubit is degenerate with respect to all clock orbits for some time $t=\tau'<\tau$ it is preferable to stop the engine there and ``run the engine in neutral'' for an additional time $\tau-\tau'$ to get back to the original setup.

For the spin-clock we can explicitly evaluate the quantity $\Gamma_{m'm}$ defined in eq. (\ref{eq:Gamma}). Starting from the definition we have
\begin{widetext}
\begin{eqnarray}\label{eq:Gamma2}
\Gamma_{m'm}(t,dt) &=& \left | \braket{m'(t+dt) | U_+(dt) | m(t)} \right|^2 \nonumber \\
&=& \left | \braket{m' | e^{iH_-(t+dt)}e^{-iH_+(dt)}e^{-iH_-(t)} | m} \right|^2 \nonumber \\
\end{eqnarray}
which looks very similar to the Wigner D-matrix \cite{Sakurai2010}
\begin{eqnarray}\label{eq:DMatrix}
D_{m'm}(\alpha,\beta,\gamma) &:=& \braket{m_z' | e^{-iL_z\alpha}e^{-iL_y\beta}e^{-iL_z\gamma} | m_z} \nonumber \\
&=& e^{-i(m' \alpha + m\gamma)} d_{m' m}(\beta)
\end{eqnarray}
where the Wigner d-matrix $d_{m'm}(\beta)$ is defined as
\begin{eqnarray}\label{eq:dMatrix}
d^l_{m'm}(\beta) = \sqrt{\frac{(l+m)!(l-m)!}{(l+m')!(l-m')!}}\sum_{\sigma=0}^{l-m'} {l+m' \choose l+m-\sigma}{l-m' \choose \sigma}(-1)^{l-m'-\sigma}(\sin\frac{\beta}{2})^{2l-m-m'-2\sigma}(\cos\frac{\beta}{2})^{m+m'+2\sigma}.
\end{eqnarray}
In (\ref{eq:DMatrix}) the states $\ket{m_z}$ and $\ket{m_z'}$ are eigenstates of the $L_z$ operator. However, in eq. (\ref{eq:Gamma2}) the states are eigenstates of $C$, not $H_-$, we thus have to introduce two identity decomposition in the $H_-$ basis $\{\ket{k_-}\}$ to make use of eq. (\ref{eq:DMatrix}). We have
\begin{eqnarray}\label{eq:Gamma3}
\Gamma_{m'm}(t,dt) &=& \left | \sum_{k_-,k_-' = -l}^l \braket{m'|k_-'}  \braket{k_-' | e^{iH_-(t+dt)}e^{-iH_+(dt)}e^{-iH_-(t)} | k_-} \braket{k_-|m} \right|^2 \nonumber \\
&=& \left | \sum_{k,k' = -l}^l D_{k'k}\left(-(t+dt),dt,t\right)\braket{m'|k_-'} \braket{k_-|m} \right|^2 \nonumber \\
&=& \left | \sum_{k,k' = -l}^l e^{-i(kt - k'(t+dt))}d_{k'k}(dt)\braket{m'|k_-'} \braket{k_-|m} \right|^2.
\end{eqnarray}
Finally, noting that $\ket{m}=e^{-iH_+\pi/2}\ket{m_-}$ and using the D-matrix result eq. (\ref{eq:DMatrix}) again we have $\braket{k_-|m} = \braket{k_- | e^{-iH_+\pi/2} | m_-}= D_{km}(0,\frac{\pi}{2},0) = d_{km}(\frac{\pi}{2})$ and similarly $\braket{m'|k_-'} = d_{m'k'}(-\frac{\pi}{2})$. Hence we arrive at 
\begin{eqnarray}\label{eq:Gamma3}
\Gamma_{m'm}(t,dt) &=& \left | \sum_{k,k' = -l}^l e^{-i(kt - k'(t+dt))}d_{k'k}(dt)d_{m'k'}(-\frac{\pi}{2})d_{km}(\frac{\pi}{2}) \right|^2.
\end{eqnarray}
\end{widetext}
This might not look more illuminating than the original expression, but the d-matrices are well known functions and can easily be evaluated computationally, allowing us to explicitly calculate results for the spin-clock example. We used this expression to generate the results shown in Figures \ref{fig:SpinWork} and \ref{fig:WorkContour}. All other results follow by simply substituting this result into the relevant expressions.

Let us conclude the analysis of the spin-clock example with the work output in the Zeno limit. Starting from $\Delta(t) = l\sin t$ we have
\begin{equation}
dF(t) = p_1(t) l \cos(t) dt
\end{equation}
which upon integration from $\frac{\pi}{2} \leq t \leq \pi$ yields
\begin{equation}
W_{Zeno} = kT\left(\log2 - \log(1+e^{-\beta l})\right).
\end{equation}
This is the maximum work that can be extracted from a pure qubit if one is limited to utilise a spin-$l$ system as a clock/machine and a thermal reservoir at inverse temperature $\beta = 1/kT$. We see that the the semi-classical $W = kT\log2$ can only be recovered for infinitely large clocks $l\rightarrow\infty$ (or for zero temperature $\beta\rightarrow\infty$).

\section{Simulating different thermalisation regimes}\label{app:therm}
As noted in the main text, to obtain the analytic results we have to make the assumption that the thermalisation of the qubit takes place at the beginning of each unit protocol, right after the measurement of the preceding unit protocol. Thus evolution (\ref{step2}) and thermalisation (\ref{step1}) are in a sense non-interacting, separated by the measurement (\ref{step3}). In this section we present simulation results that do not rely on this assumption but instead model non-trivial interactions between thermalisation and evolution, and hence show that the approximation is qualitatively robust in all the regimes considered.

\begin{figure}[tb]
\begin{center}
\includegraphics[width = 0.9\columnwidth]{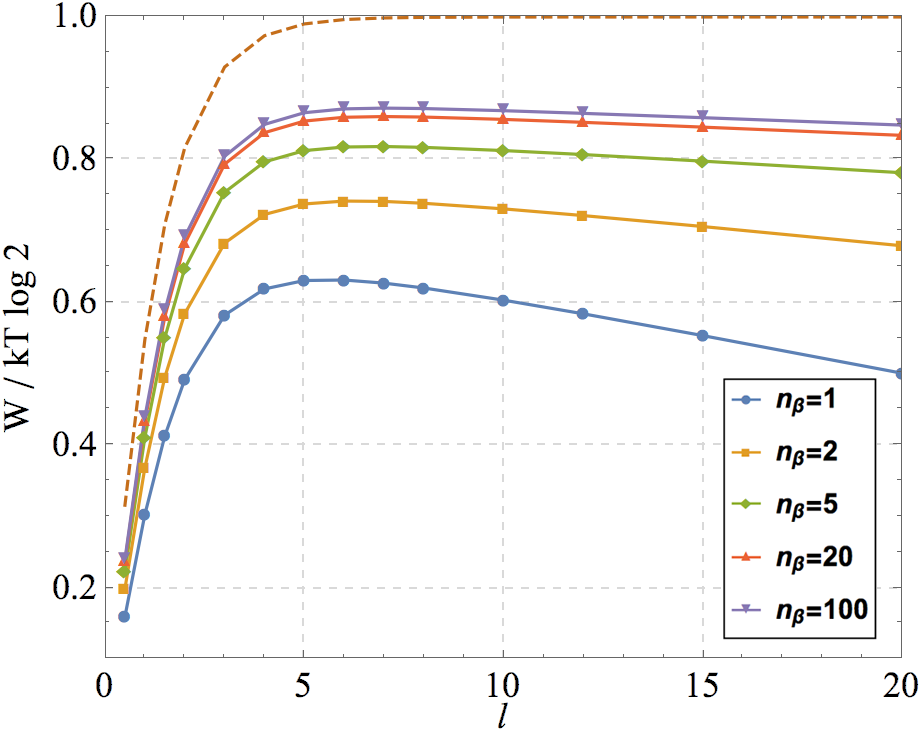}
\caption{Ideal work output of a machine with a spin-$l$ clock and stabilisation interval $dt=0.05$, plotted against clock size $l$ for different numbers of thermalisation $n_{\beta}$ per unit protocol. The $n_{\beta}=1$ result is in exact agreement with the analytic result for $dt=0.05$ in Fig. \ref{fig:SpinWork}. The higher the value of $n_{\beta}$, the more continuous the state transformation of the qubit, i.e. the more quasi-static the protocol, leading to greater work output. The dashed line shows the Zeno limit.}
\label{fig:WorkSim1}
\end{center}
\end{figure}

\begin{figure}[tb]
\begin{center}
\includegraphics[width = 0.9\columnwidth]{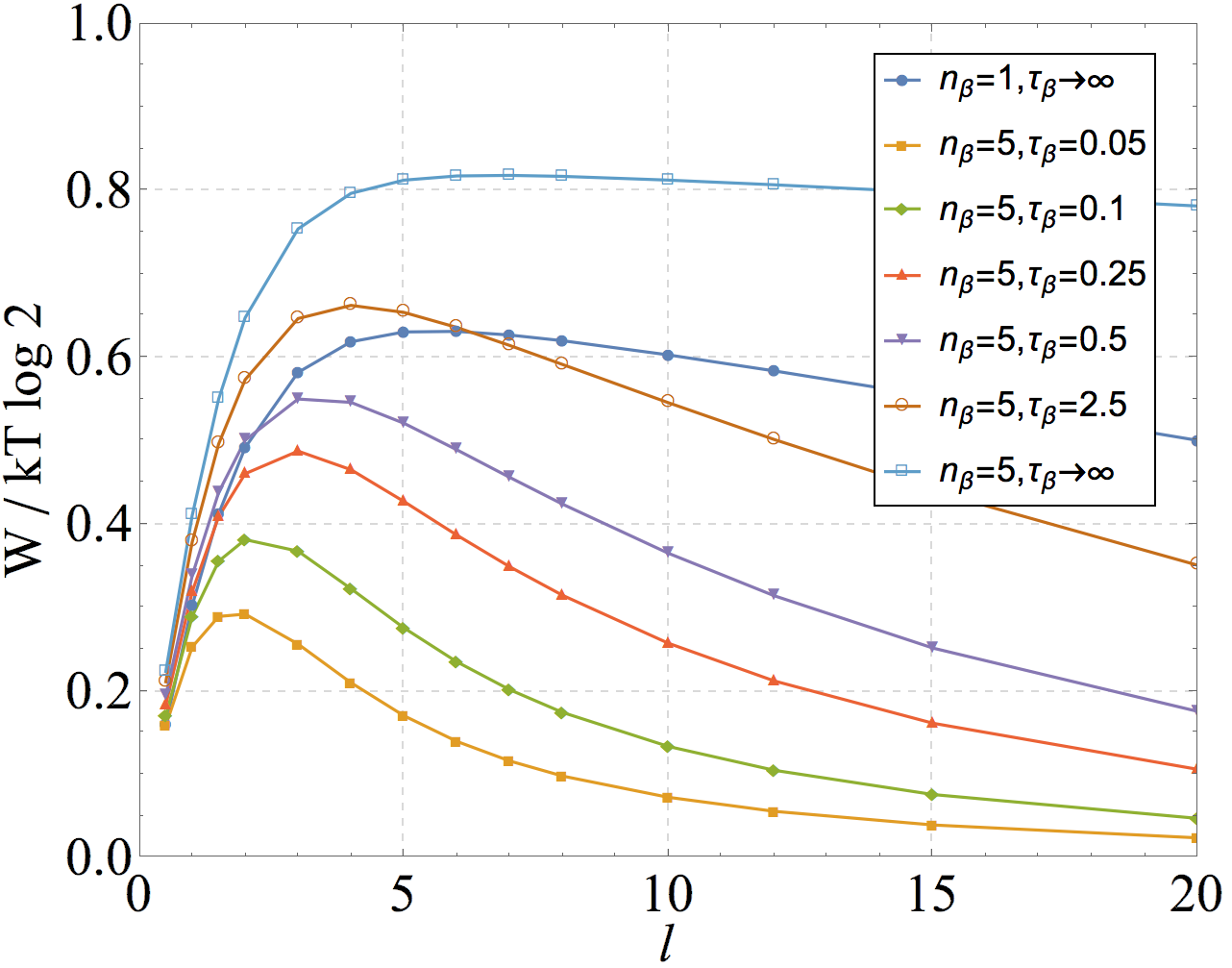}
\caption{Ideal work output of a machine with a spin-$l$ clock and stabilisation interval $dt=0.05$, plotted against clock size $l$ for $n_{\beta}=5$ finite rate equilibrations per unit protocol with different effective equilibration times $\tau_{\beta}$. The analytic result with $n_{\beta}=1$ and infinite equilibration rate $\tau_{\beta}$ is shown for comparison. Faster equilibration (larger $\tau_{\beta}$) implies an evolution of the qubit closer to equilibrium and thus larger work output. Note that for large $l$ the qubit's upper level drops faster, so that the qubit is further from equilibrium for the same equilibration rate, leading to reduced work output for large $l$.}
\label{fig:WorkSim2}
\end{center}
\end{figure}

We consider two ways of avoiding the assumption. In the first one, we simply split each unit protocol further into $n_{\beta}$ sub-units of duration $dt' = dt/n_{\beta}$, each consisting of thermalisation of the qubit followed by free evolution of the joint system for duration $dt'$. The measurement is still only performed once per UP, at the end. Note that whereas in the main text the qubit always thermalises with respect to the local Hamiltonian induced by the reference clock orbit $\chi(t)$, the thermalisations in between $t$ and $t+dt$ are with respect to the local Hamiltonian induced by the clocks momentary state, which in general deviates from the reference orbit between measurements. The more sub-units $n_{\beta}$ we consider, the more quasi-static the process becomes as the qubit more and more smoothly transitions from one thermal state to the next. Due to the immediate backaction into the clock by the joint evolution, this can be seen as the machine scanning the thermal distribution with a higher fidelity. In the limit $n_{\beta}\rightarrow\infty$ the process becomes a quasi-static equilibrium process and the work output is maximised. The results for the optimal scenario in which each measurement succeeds are shown in Fig. \ref{fig:WorkSim1} for a spin-$l$ clock with stabilisation interval $dt=0.05$ for different numbers of sub-units $n_{\beta}$. Note that the analytic result (which is equivalent to $n_{\beta}=1$ in the simulations) is not necessarily more or less realistic than the results for higher $n_{\beta}>1$, but can be seen as a non-equilibrium result similar to a finite thermalisation time of the qubit.
\\
To consider an even more realistic model of non-equilibrium behaviour, we can further introduce the notion that during each thermalisation stage the qubit is not instantaneously transformed into a Gibbs state, but instead undergoes an equilibration with a bosonic bath at a finite rate, evolving according to a standard master equation \cite{Carmichael1993}. The mean bosonic occupation number $\bar{n}$ is given by
\begin{equation}
\bar{n} = \frac{1}{e^{\beta\omega}-1}
\end{equation}
for a mode of frequency $\omega$, and we assume that at any time $t$ the qubit only couples to the mode which it is in resonance with, i.e. for which $\omega=\Delta(t)$. We further define the clock's states with respect to the qubit's $\ket{\psi}$ and $\ket{\overline{\psi}}$ states as
\begin{eqnarray}
\rho_M^{\psi} := \braket{ \psi | \rho_{SM} | \psi} \\
\rho_M^{\overline{\psi}} := \braket{ \overline{\psi} | \rho_{SM} | \overline{\psi}}
\end{eqnarray}
respectively. Using this notation it can be shown that equilibration of the qubit with a bosonic bath for an effective duration $\tau_{\beta}$ takes the joint qubit-machine state $\rho_{SM}$ to
\begin{eqnarray}
\rho_{SM}' &\propto& \ket{\psi}\bra{\psi} \otimes \left(C_{\psi\rightarrow\psi}\rho_M^{\psi} + C_{\overline{\psi}\rightarrow\psi}\rho_M^{\overline{\psi}}\right) \nonumber \\
&+& \ket{\overline{\psi}}\bra{\overline{\psi}} \otimes \left(C_{\psi\rightarrow\overline{\psi}}\rho_M^{\psi} + C_{\overline{\psi}\rightarrow\overline{\psi}}\rho_M^{\overline{\psi}}\right),
\end{eqnarray}
where the $C_{x\rightarrow y}$ transition coefficients are given by
\begin{eqnarray}
C_{\psi\rightarrow\psi} &=& \frac{e^{-(2\bar{n}+1)\tau_{\beta}}\bar{n} +\bar{n} + 1}{2\bar{n} + 1} \\
C_{\overline{\psi}\rightarrow\psi} &=& -\frac{e^{-(2\bar{n}+1)\tau_{\beta}}(\bar{n}+1) - (\bar{n} + 1)}{2\bar{n} + 1}\\
C_{\psi\rightarrow\overline{\psi}} &=& -\frac{e^{-(2\bar{n}+1)\tau_{\beta}}(\bar{n}-1)}{2\bar{n} + 1}\\
C_{\overline{\psi}\rightarrow\overline{\psi}} &=& \frac{e^{-(2\bar{n}+1)\tau_{\beta}}(\bar{n}+1)+\bar{n}}{2\bar{n} + 1}
\end{eqnarray}
In the limit $\tau_{\beta}\rightarrow\infty$ this model of equilibration corresponds to the instantaneous transformation to the Gibbs state considered above, but for finite $\tau_{\beta}$ the reduced state of the qubit will in general ``lag behind'' the Gibbs state. Using this finite time equilibration in combination with the previous notion of breaking each UP into multiple sub-units of equilibration and evolution before the final measurement allows us to model very realistic non-equilibrium behaviour. Figure \ref{fig:WorkSim2} shows the results for the spin-clock model for $n_{\beta}=5$ equilibration events and different effective thermalisation times $\tau_{\beta}$.
The analytic result corresponding to $n_{\beta}=1$ and $\tau_{\beta}\rightarrow\infty$ is also shown for comparison. 

We see that for very large $l$ the work output of the machine is drastically reduced and approaches zero in the limit of large $l$. This is due to the fact that in this specific model the level splitting of the qubit is $\Delta(t) = l\sin t$, and thus the larger $l$, the faster the level splitting changes, requiring longer effective equilibration times $\tau_{\beta}$ to keep the qubit close to its respective Gibbs state. This is in some sense equivalent to the notion encountered in the analytic results in the main text that a larger clock (i.e. faster change in level splitting) requires the machine to sample the qubit's thermal distribution at a higher rate in order to get a good work output. We again clearly see the tradeoff between maximising work output and maximising power. If we want to achieve an optimal work output we have to slow down the system dynamics (which in effect increases the ratio of effective equilibration time $\tau_{\beta}$ to the change in level splitting $\frac{d\Delta}{dt}$), which in turn reduces the power output of our engine. Conversely, increasing the power by increasing the rate of system dynamics we end up further away from the Zeno limit where we can convert the entire free energy difference of the qubit into work, thus sacrificing potential work output.
Even though the exact quantitative result strongly depends on the model parameters, we see from the results of this section that the analytic result based on the assumption of a single instantaneous thermalisation during each UP qualitatively contains all the core features and even quantitatively accurately captures the results for certain realistic thermalisation regimes of non-equilibrium dynamics. Particularly as we approach the Zeno limit all results exactly converge, if we assume a very strong coupling between qubit and bath, such that the qubit always remains (approximately) in thermal equilibrium.

\section{Fuelling the engine with mixed states} \label{app:mixed}
In this section we consider a slight variation of the selective engine mode discussed in the main text and show how the engine behaves if instead of pure states we try and fuel it with (partially) mixed states. If we consider the machine as a black-box system which takes some state $\rho_S$ as an input, generates work, and outputs a new state $\rho_S^*$, we can say that if given the ideal input  $\rho_S=\ket{\psi}\bra{\psi}$ for which the engine was designed, we get the state
\begin{equation}
\rho_S^* = q\frac{\mathbb{1}}{2} + (1-q)\ket{\psi}\bra{\psi}
\end{equation}
as the output, where $q=\prod_{n=1}^N p_d(\tilde{\tau}+n dt)$ is the probability that all $N$ unit protocols succeed an we finish the cycle with a maximally mixed qubit in the state $\mathbb{1}/2$, and $(1-q)$ is the probability that some UP fails and we perform the feedback process which returns the qubit to the original state $\ket{\psi}\bra{\psi}$. The work output associated with this state transformation $\rho_S\rightarrow\rho_S^*$ is $\braket{W}$ eq. (\ref{WAverageGeneral}). The machine itself is unchanged and will by the design of the protocol always be in the state $\chi(t)$ after outputting $\rho_S^*$, thus effectively acting as a catalyst.

But we can also ask what happens if instead of inputting the pure state $\rho_S$ into the engine, we try and feed the engine its own output state, the mixed $\rho_S^*$. We can rewrite the state as
\begin{equation}
\rho_S^* = (1-\frac{q}{2})\ket{\psi}\bra{\psi} + \frac{q}{2}\ket{\overline{\psi}}\bra{\overline{\psi}}.
\end{equation}
If we feed this state into the engine, note that the initial stage for times $0\leq t \leq \tilde{\tau}$ is now not just the trivial level splitting induction in the qubit anymore, but also leads to a deviation of the qubit from it's clock orbit. Thus we need to introduce one additional measurement before the first thermalisation to project the clock back onto its reference orbit\footnote{Note that the introduction of this additional measurement in the original protocol, which assumes pure state inputs, would be trivial since in that scenario the state commutes with the measurement.}. The probability $P^{(1)}$ of this measurement failing in projecting the clock back onto $\chi(t)$ is given by
\begin{eqnarray}
P^{(1)} &=& \frac{q}{2}\sum_{m=1}^{d-1} \left| \braket{m(\tilde{\tau}) | U_+(\tilde{\tau}) | d} \right|^2 \\
&=& \frac{q}{2}\sum_{m=1}^{d-1} \Gamma_{md}(0,\tilde{\tau})\\
&=& \frac{q}{2}\left(1- \Gamma_{dd}(0,\tilde{\tau})\right).
\end{eqnarray}
If this event occurs, the engine immediately starts a feedback process returning the qubit in the $\ket{\psi}\bra{\psi}$ at the end of the cycle. Otherwise, since the qubit is now at time $t=\tilde{\tau}$ back on its reference orbit and the qubit gets thermalised just as if it would have on input of the ideal state $\ket{\psi}\bra{\psi}$, the machine proceeds for the remainder of the cycle $\tilde{\tau} < t \leq\tau$ just as in the original protocol. As noted above, this second part, containing the actual work extraction, has a misfire and feedback event with a probability
\begin{equation}
P^{(2)} = 1 - \prod_{n=1}^N p_d(\tilde{\tau}+n dt).
\end{equation}
Putting both parts together, the chance of the machine experiencing feedback entering the feedback procedure at any point and thus returning the pure state $\ket{\psi}$ is
\begin{equation}
p(\ket{\psi}) = P^{(1)} + (1-P^{(1)})P^{(2)},
\end{equation}
whereas the engine completes the full cycle and outputs the maximally mixed state $\mathbb{1}/2$ with probability
\begin{equation}
p(\frac{\mathbb{1}}{2}) = (1-P^{(1)})(1-P^{(2)}).
\end{equation}
This allows us to ask for which $q=q^*$ the machine outputs the same state that it got as its input. This condition is met when $q^* = p(\frac{\mathbb{1}}{2})$, i.e. for
\begin{equation}
q^* = \left(1-\frac{q^*}{2}\left(1- \Gamma_{dd}(0,\tilde{\tau}) \right) \right)\prod_{n=1}^N p_d(\tilde{\tau}+n dt).
\end{equation}
In general this quantity depends on the specific machine, but we can explicitly evaluate it in the case of the spin clock in the infinite size limit $d\rightarrow\infty$. In this case we have $ \Gamma_{dd}(0,\tilde{\tau})\rightarrow0$ and $\prod_{n=1}^N p_d(\tilde{\tau}+n dt) \rightarrow 1$ such that after rearranging we find $q^* = \frac{2}{3}$.

One might intuitively expect that if the engine returns the same state that it got as an input, that at best it has a zero net work output. By explicitly calculating the relevant expressions it can easily be shown that for the stationary state with $q=2/3$ in the limit of a machine with $d\rightarrow\infty$, the energy transferred to the measurement apparatus is $\Delta E = \frac{2}{3} kT \log 2$, whereas the resetting cost of the memory is $W_{reset} \ge kT (\log 3 - \frac{2}{3} \log 2)$ such that $\Delta E-W_{reset}=kT(\frac{4}{3}\log2 - \log3) < 0$. Thus the net work output in this scenario is strictly negative. More realistic machines with finite $d$ have even lower work output. This result should not be very surprising, since a state a state with $q=2/3$ is closer to being maximally than being pure.

Instead of asking which state is stationary under the action of the machine, we can also ask which state leads to a zero net work output, such that all states more pure than this state would result in positive work. We know that such a state has to be less mixed than the stationary state, i.e. have $q<q^*$. The exact value will again strongly depend on the specific clock/machine used, but we can once more consider the classical equilibrium limit of an infinitely large clock $d\rightarrow\infty$ and the qubit being kept in equilibrium with the bath. Assuming the engine gets that far, the actual work extraction stage in this limit always succeeds, outputting an amount $kT\log2$ of work. Thus it all comes down to the probability $P^{(1)}$ of the first stabilising measurement at $t=\tilde{\tau}$ failing or succeeding. The measurement itself can easily be shown to induce a zero energy change on average (although each individual measurement result has different energy flows associated with it). Hence the total average energy transferred by the engine is $\Delta E = (1-P^{(1)})kT\log2$ which in this specific limit is equivalent to $\Delta E = (1-\frac{q}{2})kT\log2$. The memory erasure cost during the actual work extraction stage $\tilde{\tau} < t \leq \tau$ vanishes in this limit, so the only erasure cost required is the one of the initial measurement at $t=\tilde{\tau}$ which is given by $W_{reset} = kT S(\{P^{(1)},1-P^{(1)}\})= kT S(\{\frac{q}{2},1-\frac{q}{2}\})$. Hence the state with zero energy output has $q=q'$ which satisfies
\begin{equation}
(1-\frac{q'}{2})\log2 = -\frac{q'}{2}\log\frac{q'}{2} - (1-\frac{q'}{2})\log(1-\frac{q'}{2}).
\end{equation}
This equation can be solved numerically to yield $q'\approx0.454$. For any $q<q'$ the machine has a net positive work output, whereas for more mixed states with $q>q'$ the work output is negative. For realistic machines away from the infinite $d$ and perfect thermalisation limit we require even smaller $q$ (i.e. more pure states) for a positive work output.

\end{document}